# Smart and Efficient IoT-Based Irrigation System Design: Utilizing a Hybrid Agent-Based and System Dynamics Approach

Taha Ahmadi Pargo[a], Mohsen Akbarpour Shirazi[a,], Dawud Fadai[a]

a: Department of Industrial Engineering and Management Systems, Amirkabir University of Technology, 424 Hafez Ave., Tehran 15875-4413, Iran

**Abstract**
Regarding problems like reduced precipitation and an increase in population, water resource scarcity has become one of the most critical problems in modern-day societies, as a consequence, there is a shortage of available water resources for irrigation in arid and semi-arid countries. On the other hand, it is possible to utilize modern technologies to control irrigation and reduce water loss. One of these technologies is the Internet of Things (IoT). Despite the possibility of using the IoT in irrigation control systems, there are complexities in designing such systems. Considering this issue, it is possible to use agent-oriented software engineering (AOSE) methodologies to design complex cyber-physical systems such as IoT-based systems. In this research, a smart irrigation system is designed based on Prometheus AOSE methodology, to reduce water loss by maintaining soil moisture in a suitable interval. The designed system comprises sensors, a central agent, and irrigation nodes. These agents follow defined rules to maintain soil moisture at a desired level cooperatively. For system simulation, a hybrid agent-based and system dynamics model was designed. In this hybrid model, soil moisture dynamics were modeled based on the system dynamics approach. The proposed model, was implemented in AnyLogic computer simulation software. Utilizing the simulation model, irrigation rules were examined. The system's functionality in automatic irrigation mode was tested based on a 256-run, fractional factorial design, and the effects of important factors such as soil properties on total irrigated water and total operation time were analyzed. Based on the tests, the system consistently irrigated nearly optimal water amounts in all tests. Moreover, the results were also used to minimize the system's energy consumption by reducing the system's operational time.

1. **Introduction**

Since the beginning of the 21st century, humans have witnessed several technological advancements. Nowadays, it is possible to see smart devices everywhere. These sophisticated devices shape a smart network, which is called an Internet of Things (IoT) based network, and can both communicate with other devices and perform tasks autonomously [1]. Smart medical devices and smart clothes are two examples of applications of this technology.

As one of the key objectives of the 4th Industrial Revolution is the automation of processes [2], engineers have endeavored to utilize the IoT technology within many fields. IoT-based Sophisticated systems technology can collect real-time data from their surrounding environment and perform their actions based on the available data. Additionally, in recent years, the production cost of microcontrollers and sensors has become more economically friendly. Due to these advantages, it is possible to solve many engineering problems by employing these systems in various fields, from industry to agriculture.

One field in which engineers have endeavored to employ new technologies is agriculture [3]. Engineers have developed IoT solutions for diverse agricultural domains, including precision farming, autonomous harvesting, and field condition monitoring. A key challenge for which engineers have sought solutions is irrigation. Arid and semi-arid regions, such as those in the Middle East and Africa, face severe agricultural challenges [4]. Due to the significant impact of the agriculture on economic, environmental, and social aspects of people's lives, many researchers have been motivated to provide solutions for agricultural challenges. Many scientists and engineers have attempted to address agricultural challenges with novel technologies such as machine learning and IoT. IoT-based systems can potentially reduce costs by minimizing the waste of resources [5]. many water-scarce countries, often developing nations, may rely on inefficient traditional irrigation methods [6]. In recent years, engineers and scientists have strived to provide more efficient, cost-effective IoT-based irrigation solutions.

Most traditional irrigation methods waste a huge amount of water in the long term. In order to irrigate optimally, farmers could irrigate their lands under two conditions [7]:
1) Irrigate as much water as it transpires from the plants and evaporates from the field
2) Maintain the soil moisture below the level that the moisture percolate to a deeper depth of soil, and above the level that plant roots cannot absorb the moisture.

By adhering to these two conditions, the irrigation amount should meet its optimal value. complying with both conditions simultaneously is challenging. Conversely, IoT-based solutions offer a viable approach to meeting these requirements [3].

Despite the benefits of IoT-based systems, designers have faced several challenges in designing these systems due to their complexity. One suitable approach to designing IoT-based networks is the agent-based approach. The concept of an agent shares numerous similarities with IoT-based devices. The word "agent" means a sophisticated and autonomous entity, which can communicate, influence, and get influenced by other agents and the environment. These definitions share broad similarities with the definitions of smart objects in IoT-based networks, making the agent-based approach well-suited for modeling IoT networks. With this approach, it is possible to design and model the system and simulate it before implementation in real life [8]. By modeling and simulating IoT-based systems, it is possible to check their functionality before implementation.

Additionally, many agent-oriented software engineering (AOSE) methodologies have been created to help designers model the systems with an agent-based approach [9]. By utilizing these methodologies, it is possible to model the IoT nodes and the environment. In this study, we aim to design a smart IoT-based irrigation system using a hybrid agent-based and system dynamics approach. The AOSE methodology "Prometheus" was employed to design the system. For this system, real-time, data-driven, and robust rules are designed. Also, the soil moisture dynamics, which is part of the system's surrounding environment, were modeled based on the system dynamics approach, which provides flexible tools to model dynamic systems. In this hybrid model, the agent-based system and the soil moisture dynamics affect each other directly. The designed system can operate both automatically and manually, based on the choice of the operator. Furthermore, considering the system's impact on surrounding soil moisture, a soil moisture dynamics model was constructed based on system dynamics principles. This model was integrated into the agent-based model as the environment, and can both influence and get influenced by system performance.

Two main goals were followed in designing this system: to irrigate optimally and operate with the minimum energy consumption so the system can run efficiently under battery limitation. To analyze the functionality of the designed system, a computer simulation model was constructed based on the designed model. The simulation model was used to test the system's behavior in different scenarios. Moreover, it is possible to examine the simulation outcomes by utilizing design of experiments techniques [10]. The system was tested in a fractional $2^{13-5}$ experimental design, to investigate its performance in different soil types. Additionally, the system's energy consumption was reduced by minimizing system's operational time by setting the adjustable parameters to their proper values. The response surface methodology was utilized to find the proper values of the adjustable parameters.

The main goals of this paper are:
1) Model a smart IoT-based irrigation system, based on the agent-based approach
2) Design smart rules to make the system operate optimally, and test the systems functionality, utilizing the simulation model
3) Decrease the system's operation time in order to reduce system's energy consumption, utilizing the simulation model

To achieve these goals, the literature is first reviewed. In the next step methodologies and the related definitions are introduced. Then, the hybrid agent-based and system dynamics model is designed to simulate the system and the soil moisture, which influence each other. In the following, the system is designed based on Prometheus methodology. The model is tested to check if it is working properly. Then the system is tested in various scenarios to check if it is working as desired and to find suitable parameter values to minimize the system's operating time. Finally, the conclusions are made.

## 2. Literature Review

Due to the fact that there is a lot of research in both the fields of IoT-based irrigation systems and also the design of IoT networks using agent-based modeling, in this section the articles that form the basis of these two issues are reviewed.

*2.1. IoT-based irrigation systems*

In recent years, many articles have investigated smart irrigation systems based on the IoT technology, including communication technologies, processors, and sensors. Many of these articles have only designed systems that make irrigation decisions based on soil moisture. However, some articles have also used the approaches of fuzzy rules, machine learning and data analysis in this regard.

Gutiérrez et al. designed an irrigation system using Wi-Fi technologies, a GPRS module, and soil moisture sensors. In their designed system, Irrigation decisions are made in this system by an algorithm based on temperature and soil moisture [11]. The system designed by Bennis et al. is also designed based on establishing communication on the basis of Wi-Fi networks. Their proposed system, in addition to soil moisture and temperature, considers environmental pressure to be effective in decision-making and uses the information collected by the pressure sensor to make decisions [12]. Ragheb et al. also developed an irrigation system based on IoT technology that uses information obtained from temperature, soil moisture, and rain sensors. They also created a web application with a user interface for their model [13]. Goap et al. have also used cloud computing and machine learning to design their proposed system. Their designed system in addition to information like air temperature, air humidity, soil temperature, and soil moisture, also uses an ultraviolet sensor to perform calculations [14].

In addition to this research, scientists and engineers have also utilized modern technologies related to machine learning to enhance irrigation systems. In their proposed system, Nawandar and Satpute trained a neural network to make irrigation decisions. Another modern soft computing technique is utilizing the fuzzy rules [15]. Krishnan et al. used fuzzy rules to control the irrigation system. In this system, first, the parameters are converted into fuzzy values, and then the decision is made based on the fuzzy rules [16].

As a branch of machine learning, one technology that can enhance irrigation is image processing. Kwok and Sun also designed an intelligent irrigation system that has the ability to recognize the plant type utilizing image processing techniques [17]. The irrigation system designed by Behzadipour et al. is a system based on prediction models that work with temperature, humidity, light data, and information obtained from image processing of tree leaves [18]. Rao et al. also developed a system that, in addition to irrigation, by using image processing techniques and information from sensors, it possible to detect plant diseases and minerals in water [19].

Finally, it is important to mention that recently researchers have used simulation techniques to design irrigation systems. Gomez Alves et al. introduced an irrigation system based on IoT technology. Before implementing their system, they used discrete event simulation to check the performance of their system [20]. In another study published in 2024, Manocha et al. developed an approach for the optimal control of smart irrigation based on the IoT using a digital twin, which also has the ability to simulate the system [21].

*2.2. Utilizing the agent-based modeling to design the IoT-based systems*

The definition of an agent in agent-based modeling is very close to the definition of smart IoT-based devices. These devices make decisions autonomously, are influenced by the environment and can influence the environment, are in communication with each other, and can influence each other. According to these concepts, the use of agent-based methodology is considered a suitable approach for modeling systems based on the Internet of Things.

In 2016, Savaglio et al. proposed the use of an agent-based approach to model IoT-based systems [22]. The similarity in defining devices based on the Internet of Things and the agent in agent-based modeling indicates the appropriateness of using this approach for modeling systems based on the Internet of Things.

The research conducted by Kaminski et al. is also among the primary researches in this field. In this paper, researchers modeled an IoT-based network with an agent-based approach [23]. Batool and Niazi also investigated the use of an agent-based approach to model complex IoT-based networks [24]. Additionally, Pérez-Hernández et al.

developed an agent-based simulation model to design systems based on the Internet of Things [25]. Their proposed approach is used to test the performance of the designed system.

Wang et al. proposed a data communication mechanism for smart greenhouses which utilize IoT technology, based on an agent-based approach [26]. Gomes et al. designed a multi-agent system based on the IoT to intelligently monitor the energy used in homes [27]. In another study, Su and Wang investigated the impacts of uncertain information delays on distributed real-time optimal controls for building HVAC systems deployed on IoT-enabled field control networks [28].

*2.3. Research gap and contributions*

In general, various research studies have been done in both the fields of smart, IoT-based irrigation systems and the design of IoT-based systems utilizing the agent-based approach. The table 1 shows the summarized review of the related literature.

Table 1. Summarized review of the related literature

| Author(s) | Publication year | Simulation-based approach | Agent-based approach | IoT-based system design | Irrigation system | Designed to reduce energy consumption |
|---|---|---|---|---|---|---|
| Gutiérrez et al. [11] | 2013 | | | ✓ | ✓ | |
| Bennis et al. [12] | 2015 | | | ✓ | ✓ | |
| Batool and Niazi [24] | 2017 | ✓ | ✓ | ✓ | | |
| Goap et al. [14] | 2018 | | | ✓ | ✓ | |
| Kwok and Sun [17] | 2018 | | | ✓ | ✓ | |
| Kaminski et al. [23] | 2018 | ✓ | ✓ | ✓ | | |
| Pérez-Hernández et al. [25] | 2018 | ✓ | ✓ | ✓ | | |
| Gomes et al. [27] | 2018 | ✓ | ✓ | ✓ | | |
| Nawandar and Satpute [15] | 2019 | | | ✓ | ✓ | |
| Krishnan et al. [16] | 2020 | | | ✓ | ✓ | |
| Rao et al. [19] | 2020 | | | ✓ | ✓ | |
| Wang et al. [26] | 2020 | | ✓ | ✓ | | |
| Ragheb et al. [13] | 2022 | | | ✓ | ✓ | |
| Behzadipour et al. [18] | 2023 | | | ✓ | ✓ | |
| Gomez Alves et al. [20] | 2023 | ✓ | | ✓ | ✓ | |
| Su and Wang [28] | 2023 | ✓ | ✓ | ✓ | | |
| Manocha et al. [21] | 2024 | ✓ | | ✓ | ✓ | |
| Current study | | ✓ | ✓ | ✓ | ✓ | ✓ |

Although researchers have recently used tools such as machine learning in designing smart irrigation systems, many of the designed systems only use basic information such as temperature and soil moisture. Most of the research has only done irrigation based on the amount of soil moisture, while an optimal system needs to calculate the amount of water required and predict the soil moisture in such a way that irrigation is done before the plant is stressed. On the other hand, in the real world, we face various limitations. For example, the use of appropriate tools and sensors for image processing requires appropriate hardware, while the use of this hardware may not seem reasonable in terms of cost. The primary goal of this research is to calculate the amount of water needed at the right time and irrigate in such a way that water is not wasted and water stress is not applied to the plant.

On the other hand, as mentioned, researchers have recently tried to simulate irrigation systems to test the system's performance. Considering the consistency of using agent-based modeling in the design of IoT-based systems, in this research, an irrigation system is designed using this approach.

One of the key advantages of using the agent-based approach is the ability of testing the system's performance by monitoring the agents during the simulation. This feature allows it to test the system's functionality. Based on this capability, the designed model is simulated in AnyLogic software to check its performance.

In addition, based on the literature review, no articles were found that aim to minimize system energy consumption. In this study, the numerical results obtained from the simulation are used to minimize the irrigation system's energy consumption by minimizing the devices' operating time.

Finally, it is worth mentioning that soil moisture dynamics were modeled in this study. Soil moisture is an important element in irrigation decision-making that changes dynamically and depending on environmental factors. To investigate the performance of the irrigation system using simulation, soil moisture dynamics must be simulated to determine the system's performance. Given that the irrigation system decides to irrigate or not based on the soil moisture, and on the other hand, irrigation also changes the soil moisture, a hybrid approach was carried out to model the system and its surrounding environment so that the agent-based model of the system interacts with the soil moisture dynamic model, which was designed using the system dynamics approach, and the performance of the system is measured.

## 3. Methodologies

### 3.1. Agent-Based Modeling

Agent-based modeling (ABM) is a bottom-up approach used to model complex systems consisting of various elements. By employing this approach, these systems can be modeled with less complexity [29]. Simulation software such as NetLogo and AnyLogic are utilized to simulate these models [30].

Systems in ABM consist of various autonomous agents capable of making decisions in different conditions and can have their own independent policies. These agents can communicate with each other and their surrounding environment. Additionally, their decisions can impact other agents and the environment, while they themselves can be influenced by others and their surroundings, and finally, these interactions shape the systems behavior in different conditions [31].

In general, using the bottom-up approach in ABM allows systems to be designed by modeling agents and simulating them under various scenarios to understand the implications of different decisions in planning and designing these systems. Fig. 1 illustrates a simple diagram to show how agents interact each other and the environment in agent-based methodology.

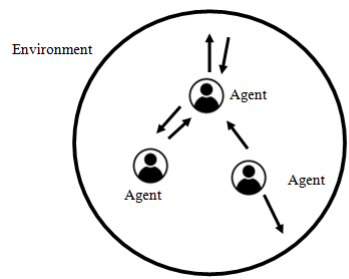

Fig. 1. A simple diagram showing how agents interact each other and the environment

Agent-based modeling has various benefits, some of which include as following:
- Ease in modeling complex systems.

- The ability to investigate the emergent behavior of system components.
- Ease in understanding complex systems.
- Possibility to examine different behaviors of various entities.
- Capability to examine the system in various scenarios.

*3.2. Prometheus Methodology*

The Prometheus methodology is an agent-oriented software Engineering methodology designed with a focus on practicality in the real world. Alongside its precision in identifying various system components, functionalities, and flexibility, it's noteworthy for its easy learning curve, especially for beginners. Fig. 2 illustrates an overview of Prometheus methodology.
In general, modeling in this methodology occurs in three phases [32]:

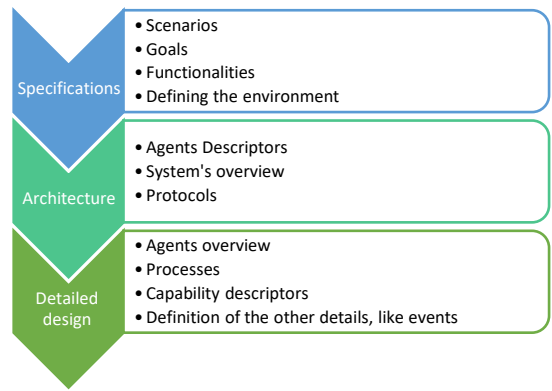

Fig. 2. Prometheus methodology overview

The first phase of the Prometheus methodology involves determining the system specifications. During this phase, it's necessary to identify the scenarios under which the system will function, including specific situations that might arise. Additionally, defining the system's goals and functionalities is crucial.

In the second phase, which focuses on the system's architecture, begins with defining different types of agents. Each type of agent may possess one or more functions. This phase also involves specifying relationships between agents and between agents and the environment. Determining how an agent's actions affect the environment and other agents is essential. Lastly, outlining the overall structure of the system where system boundaries are defined, and different types of agents are placed within this structure.

The third phase in the Prometheus methodology involves designing system details. This phase requires detailing precise behaviors of the agents. These behaviors must be explicitly specified. If agents use specific decision-making algorithms, these algorithms need to be defined in this phase. Alongside defining precise functions, existing processes wherein agents mutually influence each other must be specified. Mutual effects between agents and the environment should also be precisely defined. Lastly, the final part of the third phase in the Prometheus methodology involves specifying details regarding events and data.

*3.3. System Dynamics Modeling*

System Dynamics (SD) modeling helps us understand and simulate how complex systems change over time. It looks at how things like feedback, how much we have, how things move between these quantities, and how long it takes for changes to happen can explain how lots of different systems work. Casual loop diagrams and Stock and Flow models are usually used in order to model the complex systems [33].
It is possible to model a system by utilizing the following elements [33]:

- Stocks: Stocks are like containers that hold something, for example water in a tank. They show how much of something there is at a specific moment, and they can change over time.

- Flows: Flows are like the actions that fill or drain those containers (stocks). For instance, the entry and the output of the water from the tank. They show how things are changing in or out of those containers over time.
- Feedback loops: Feedback loops are connections between what's accumulating (stocks) and what's moving (flows) in a system. They can either boost or ease changes happening in the system. When they magnify changes, they're called positive feedback loops. If they counteract changes, they're known as negative feedback loops, which help keep things steadier.
- Delays: Time delays are like the lag or waiting time between making a change and seeing its impact elsewhere in a system. It's like when you adjust something, but it takes a bit for that change to show up in another part. This delay can make things behave in unexpected ways, causing swings or going past the intended mark before settling.

By utilizing these concepts, it is possible to model the complex systems by utilizing SD approach and test their behavior during periods of time. It is worth to use SD models in different fields like business dynamics modeling, economics, healthcare, and Industry.

*3.4. Design of experiments*

Design of experiments (DOE) is a statistical tool deployed in various types of systems and processes and can be used for multiple purposes, like variable screening, robust design, and optimization [34]. DOE techniques can be used to examine computer models and optimize the existing parameters [10].

While it is possible to use the analysis of variance technique in order to determine significant factors on a response variable, and fit regression models to predict the response variable, it is also possible to use response surface methodologies and robust parameter design techniques to set the adjustable parameters on suitable levels to optimize the results [10].

Several designs exist for conducting experiments. One series of these designs are factorial designs. The most important of these special cases is k factors, each at only two levels [10], and is called $2^k$ factorial design. By increasing the number of existing factors, the number of experimental runs increases. For example, an experiment with 10 factors with a $2^{10}$ design consists of 1024 experimental runs, which may not be cost-efficient. If the experimenter can reasonably assume that certain high-order interactions are negligible, information on the main effects and low-order interactions may be obtained by running only a fraction of the complete factorial experiment. These fractional factorial designs are among the most widely used types of designs for product and process design, process improvement, and industrial/business experimentation [10]. A resolution V design, which allows all main effects and two-factor interactions to be estimated by assuming that three-factor and higher interactions are negligible [10], is a good choice for many studies.

## 4. Definitions
*4.1. Soil properties*
- Relative volumetric soil moisture: It is called relative to the volume of water in the soil, to the total volume of the soil. Relative soil moisture may vary from zero to the saturation point of the soil [35].
- Wilting point: The wilting point is the volume of the soil that if the relative soil moisture is reduced to such an extent, the plants will not be able to absorb water from the soil and after the passage of time and the loss of the water stored in the plant, they will wither [35].
- Soil saturation point: the maximum volume percentage that may contain moisture in the soil. This percentage is usually equal to the percentage of soil porosity, and when a large volume of water enters the soil, the entire porosity in the soil is filled by water. In this case, water does not have the ability to enter the soil, and water does not enter the soil until a part of the water has penetrated to the lower layers. It is worth mentioning that part of the water that has not entered the soil can be lost as runoff. The amount of water lost is formulated using equation 1 [36]:

$$TRO = R_{coeff} \cdot TSW \tag{1}$$

In this regard, *TSW* represents the total surface water, $R_{coeff}$. the runoff constant, and *TRO* the total runoff.

- Field capacity: It means the amount of water that the soil can hold. After the soil reaches the saturation point, a part of the soil moisture that the soil cannot hold penetrates the lower layers and a part of the moisture that the soil can hold remains in the soil [35].
- Threshold point: the amount of soil moisture, if the soil moisture is less than that, the plant will experience stress in absorbing water. And the rate of sweating decreases. This point is between the wilting point and soil capacity.
- Total available water: the volume of soil moisture that is between the wilting point and soil capacity. This parameter is the difference between wilting points and soil capacity, and it is the total amount of water available to the plant because if the soil moisture is more than the soil capacity, it penetrates the lower layers [7].
- Readily available water: the volume of soil moisture that is between the threshold point and soil capacity. This amount is a fraction of the soil moisture that the plant can absorb without experiencing stress in absorbing water. The amount of water ready for access is calculated using the following equation [7].

$$RAW = P \cdot TAW \tag{2}$$

*RAW* represents the readily available water, *TAW* is the total available water and *P* is a fraction that shows how much of the total available water is ready to be accessed and varies depending on the type of plant. The value of this parameter for different plants is given in source.

In general, Fig. 3 illustrates the order of soil properties related to soil moisture.

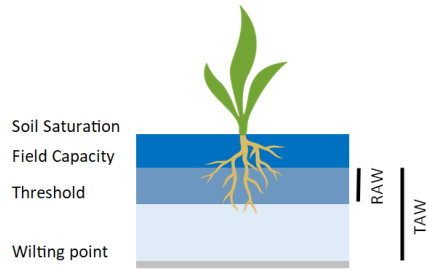

Fig. 3. Order of soil properties related to soil moisture

- Infiltration rate: it is called the speed of water entering the soil. If the soil has reached the saturation point, water does not enter the soil.
- Percolation rate: It is called the speed of water transfer in the soil where the moisture present in it is more than the soil capacity to the lower layers of the soil.

### 4.2. *Definitions related to irrigation*

Irrigation means keeping the soil moisture in conditions suitable for plant growth. If the soil moisture exceeds the field's capacity during irrigation, part of the water will percolate to deeper levels, and if the soil moisture decreases to a volume lower than the threshold point, the plants will experience stress in absorbing water; Therefore, the goal of irrigation in an ideal state is to maintain water in the interval between threshold moisture and field's capacity, in which case the water requirement of the plant is equal to the total evapotranspiration [35].

Equation 3 is used to find evaporation and transpiration of surfaces composed of different plants [7]:

$$ET = K_C \cdot ET_0 \tag{3}$$

In this regard, $ET_0$ indicates evapotranspiration and $K_C$ is the crop factor, which is different in different plants. The amount of plant factor varies during the growth of the plant and is obtained based on the current growth of the plant and the type of plant and using prepared tables (available in the source. $ET_0$ also means reference evapotranspiration, which means the amount of evaporation and transpiration available from the surface consisting of grass that is sufficiently irrigated.

In the materials published by the Food and Agriculture Organization of the United Nations, if the soil moisture is less than the threshold moisture, the amount of plant transpiration decreases linearly until it becomes zero at the wilting point. Considering that stress in water absorption reduces transpiration, the evaporation and transpiration factor is divided into two constants, one related to plant transpiration and the other related to evaporation, and equation 4 is rewritten as follows [7].

$$ET = (K_{cb} + K_e) \cdot ET_0 \tag{4}$$

$K_{cb}$ is related to plant transpiration and $K_e$ is related to water evaporation. Now, in the condition that the soil moisture is less than the threshold moisture point, the coefficient $K_s$ is also multiplied by the constant associated with plant transpiration, and in this condition, the rate of evaporation and transpiration can be calculated with the following relationship [7]:

$$ET = (K_s \cdot K_{cb} + K_e) \cdot ET_0 \qquad (5)$$

that when the soil moisture is less than its threshold point, the amount of $K_s$ can be calculated using the equation 5, as it is known, as the amount of soil moisture decreases, with the decrease of the amount of $K_s$, despite the plant's need for water, the plant's transpiration will be decreased until the plant's transpiration reaches zero at the wilting point [7].

$$K_s = \frac{S_a - S_{wp}}{S_{th} - S_{wp}} \qquad (6)$$

Where $S_a$ is the amount of soil moisture, $S_{wp}$ is equal to the amount of soil moisture at the wilting point, and $S_{th}$ is equal to the amount of soil moisture at the moisture threshold point.

Different relationships are also used to find the basic evaporation and transpiration rate, and according to the amount of available data, the appropriate relationship is chosen to estimate this rate. To find the rate of basic evaporation and transpiration, the following relations can be used:

**Penman-Monteith equation:** This equation has been recommended by the Food and Agriculture Organization of the United Nations as a suitable and accurate equation for finding the reference evapotranspiration [7].

$$ET_0 = \frac{0.408\Delta(R_n - G) + \gamma\left(\frac{900}{Tm + 273}\right)u_2(e_s - e_a)}{\Delta + \gamma(1 + 0.34u_2)} \qquad (7)$$

In equation 7, $ET_0$ is daily reference evapotranspiration (mm). In order to calculate hourly reference evapotranspiration, the calculated $ET_0$ should be divided by 24. $\Delta$ is rate of change of saturation specific humidity with air temperature (Pa.K$^{-1}$). $R_n$ is net irradiance (MJ.m$^{-2}$.day$^{-1}$), $G$ is ground heat flux (MJ.m$^{-2}$.day$^{-1}$), $\gamma$ is psychrometric constant ($\gamma \approx 66$ Pa.K$^{-1}$), $u_2$ is Wind speed at 2m height (m.s$^{-1}$), $Tm$ is Air temperature at 2m (°C), $e_s$ is mean saturation vapor pressure and (kPa), $e_a$ is mean ambient vapor pressure (kPa).

According to the inputs of the Penman-Monteith relationship, we need a lot of data to obtain reference evapotrasnpiration, which has caused other equations to be written to estimate the amount of basic evaporation and transpiration, and according to the amount of available information, the appropriate relationship can be used.

**Blaney-Criddle equation:**

$$ET_0 = \rho(0.46T_m + 8) \qquad (8)$$

In this regard, $ET_0$ is the amount of evaporation and transpiration (mm.day$^{-1}$), $T_m$ is the average temperature (°C) and $\rho$ is the value that is obtained using predefined tables, and according to the latitude of the place [37].

In general, the most accurate values are obtained using the Penman-Monteith relationship, which are very accurate in calculating the actual rate of reference evapotranspiration. The Blaney-Crridle equation is also recommended when meteorological data is scarce [38].

## 5. Designing the structure of a Smart Irrigation System

In this step, considering the stages described regarding the Prometheus methodology, we proceed to explain the modeling process. This section includes three main phases: system specifications, system architecture, and system details, which will be elaborated on in detail below.

*5.1. System Specifications*

This phase consists of three subsections:
• Scenarios
• Goals
• Functionalities

In essence, this phase deals with outlining the generalities of the model, which will be further specified with the necessary details related to the irrigation issue.

**Goals:** The main goal of this system is to reduce the irrigated water, while the crops must grow regularly. In fact, the exact definition of the main goal of this system in automatic mode is to maintain the soil moisture in the interval between the threshold moisture point and the soil capacity, and irrigation is the amount of evaporation and transpiration. Moreover, the system should operate with the minimum energy consumption.

**Scenarios:** In this model, several primary factors impact the agents' performance. Rainfall, decreasing available water levels, and miscalculations in estimating required water concerning plant, soil, and weather conditions constitute these characteristics.

Scenarios include situations where the system's behavior might change based on the predefined rules. These scenarios are created based on the following:
- Normal situation: A state where rain is not falling, and the soil moisture is between the wilting point and the threshold point
- Unpredicted reduction of soil moisture below the wilting point: If the system is in automatic mode, and irrigation is also interrupted, in this case irrigation will be done immediately.
- Unpredicted increase of soil moisture beyond the soil capacity: If the system is in automatic mode and irrigation is also being done, it will be stopped immediately.
- Unpredicted rain: In this case, if irrigation is in progress, it will be interrupted.
- Errors in calculations and parameter measurements: If, at the end of the day, the irrigated water amount differs significantly from the predicted amount, there might be errors in measurements and calculations. Also, if soil moisture increases to levels higher than field capacity or decreases to levels less than threshold point, there should be a miscalculation.
- Sensor malfunctions: If sensors report unpredictable values with a low probability, there might be a need to check the sensors.

**Functionalities:** Functionalities involve behaviors that must occur under specific conditions for the system to be directed toward a specific goal. The system's required data should also be specified in this section.

The functions of the system are as follows:
- Waking up the system in time intervals to check the soil moisture and irrigate if needed
- Measurement of required environmental parameters by sensors
- Calculating the amount of water needed by calculating the amount of evapotranspiration based on the type of plants and environmental conditions
- The ability to make decisions about irrigation if needed
- The ability to water according to manual schedules if the user wishes
- The ability to reduce the irrigation of each section according to the expert's opinion in the case of water shortage
- Send an error or warning to the user if needed

Each functionality, utilizes the existing data to operate. The Fig. 4 shows the relations between the functionalities and the existing data.

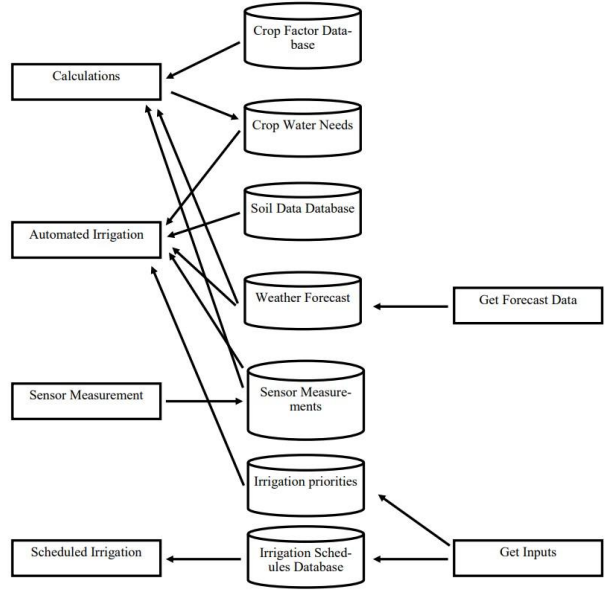

Fig. 4. Relations between the functionalities and the existing data

*5.1.1. Modeling the Environment*

The environment in this model encompasses environmental elements impacting the system's performance. Broadly, three influential factors in designing the environment have been considered. Weather, influenced by temperature and other meteorological data, rainfall can impact soil moisture, and soil condition itself. Temperature and chance of rainfall play significant rules in irrigation and they influence the controller devices in the system. Weather, UV index and data related to precipitation can be collected from online services.

Another environmental element that can be influential is soil condition. Modeling soil in this environment holds significant importance. By considering the soil type and characteristics such as water infiltration rate, percolation rate, and soil capacity, soil moisture dynamics can be modeled. In our proposed approach Soil moisture is an important factor to make the decision whether to irrigate or not, so the soil moisture influences the agents. Also, After the decision is made, the result of decision impacts the soils moisture. If agents decide to irrigate, the soil moisture will increase. Otherwise, the soil moisture will decrease due to evapotranspiration and percolation.

Soil moisture is dependent on various conditions, such as the infiltration rate, percolation rate, evaporation, and transpiration. The dynamics of soil moisture can be modeled based on a SD approach. The Fig. 5 illustrates the relation of the hybrid agent-based and SD approach to model the system and the environment.

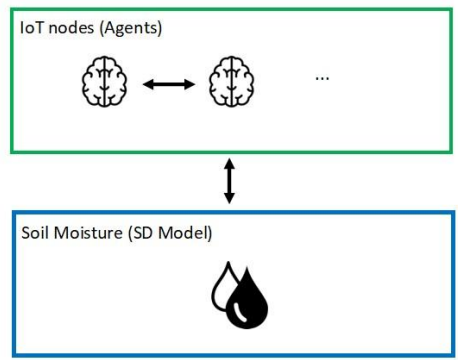

Fig. 5. Interaction of soil moisture dynamics model with existing agents

The aim of this model is to simulate the dynamics of moisture present in the soil. Considering influential factors like irrigation, rainfall occurrences, and percolation, regression equations cannot be utilized to model the soil moisture dynamics. Typically, the trend of soil moisture changes every few hours. When soil saturates and its pores fill with water, soil moisture increases to soils saturation point. At this point, water within the soil percolates to deeper soil layers, until soil moisture level decreases to fields capacity. Fields capacity is the moisture level that soil can store water in itself. As water evaporate from soil and transpire from the crops, the soil moisture still decreases. Finally, without irrigation or precipitation, soil moisture decreases to the threshold point. This situation is stressful for crops. Ideally, the moisture level should remain within a limited range between threshold point and fields capacity, allowing the plant roots to absorb it while ensuring it doesn't exceed the soil's capacity, preventing excess water from percolating down. In this ideal irrigation, total irrigated water should be very close to total evapotranspiration.

The following assumptions were made in this model:
- Water movement to the lower layers and entry into the soil is smooth and uniform.
- Soil moisture is uniformly absorbed by the soil.
- Surface evaporation is not taken into account because if the system behaves within its proper limits, the soil moisture will never reach the point where water on the soil is considered.
- Water absorption by the root is assumed to be uniform in the root area.

Model parameters, stocks and flows are listed in Table 2.

Table 2. Elements of the SD model of soil moisture dynamics

| Element | Type | Description |
|---|---|---|
| **surfaceWaterContent (mm)** | Stock | It indicates the height of water in a column to the cross-sectional area of one square milliliter on the surface. (The height of surface water that increases with irrigation.) |
| **soilWaterContent (mm)** | Stock | It indicates the height of water in a column to the cross-sectional area of one square milliliter of soil volume. It is possible to calculate the soil moisture, which is important parameter, by dividing soilWaterContent to volume of the soil. |
| **totalEVT (mm)** | Stock | It indicates the height of the total water that has evapotranspirated. |
| **totalPercolation (mm)** | Stock | The indicator of the total height of the water that has percolated to the lower layers. |
| **totalRunoff (mm)** | Stock | It indicates the amount of surface runoff. A part of ponded water leaves the surface as runoff. |
| **irrigationAndRain (mm)** | Flow | It is an indicator of water that is being added to the soil surface by irrigation and rain. |
| **Infilteration (mm/min)** | Flow | It indicates the flow of water that infiltrated into the soil. If the stored water in the soil is more than the saturation point, this flow will be equal to 0. When the irrigation and rates are lower than the maxInfiltrationRate parameter and there is no ponded water, this flow is equal to the irrigationAndRain flow, and when there is ponded water or the irrigation rate is more than the maxInfiltrationRate, this flow will be equal to the maxInfiltrationRate. In general, this flow is controlled by the infiltrationControl loop, and if the soil moisture increases beyond the critical level, this flow reaches zero. |
| **Percolation (mm/min)** | Flow | It indicates the flow of water that percolates into the lower layers. If the amount of water in the soil is less than the capacity of the soil, the water is stored in the soil and does not percolate into the underlying layers, and this flow will be equal to 0, and otherwise, this flow is equal to the percolationRate parameter. |

| Element | Type | Description |
|---|---|---|
| **Evapotranspiration (mm/min)** | Flow | Water that is evapotranspirating. This flow is modeled based on equations 3 and 4, so in the case that the soil moisture exceeds the soil moisture threshold, this value is equal to the multiplication of the reference evapotranspiration rate by the sum of the Kcb and Ke parameters. And otherwise, the Kcb is also multiplied by Ks, which can be obtained from equation 6. |
| **Runoff (mm/min)** | Flow | The amount of surface runoff. The amount of this rate is determined based on the runoff coefficient, which is based on the type of surface. This constant represents a part of the surface water that comes out as runoff. If the amount of surface water is equal to zero, the amount of this flow will also be equal to 0. |
| **irrigationRate (mm/min)** | Flow parameter | During irrigation, this rate shows how much water is irrigated per unit of time. |
| **rainRate (mm/min)** | Flow parameter | Rainfall rate |
| **MaxInfilterationRate (mm/min)** | Flow parameter | The maximum volume of water that enters the soil in one unit of time. |
| **percolationRate (mm/min)** | Flow parameter | The volume of water that leaves a unit volume of soil in a unit of time. |
| **refEvapotranspirationRate (mm/min)** | Flow parameter | Reference evapotranspiration rate. |

| Element | Type | Description |
|---|---|---|
| **Kcb** | Constant | The transpiration constant in crop coefficient |
| **Ke** | Constant | Evaporation constant in crop coefficient |
| **wiltingPoint** | Constant | It indicates the amount of soil moisture that plants are not able to absorb water from the soil. |
| **saturationCapacity** | Constant | the maximum volume percentage that may contain moisture in the soil. |
| **fieldCapacity** | Constant | It indicates the amount of soil moisture that the soil has the ability to maintain and does not enter the lower layers |
| **rootZone** | Constant | The depth of the root zone. |
| **runoffCoeff** | Constant | The amount of runoff that is determined based on the type of surface. |
| **p** | Constant | P is a fraction that shows how much of the total available water is ready to be accessed and varies depending on the type of plant. |
| **infilterationControl** | Balancing loop | This loop shows how the infiltration is controlled. If the amount of soil moisture reaches more than the saturation point of the soil, the infilteration rate will be zero. Otherwise, depending on whether the total rate of irrigation and rainfall is less than the maximum rate of water entering the soil or not, this flow increases. |

Moreover, the Fig. 6 illustrates the stock and flow diagram of soil moisture dynamics model.

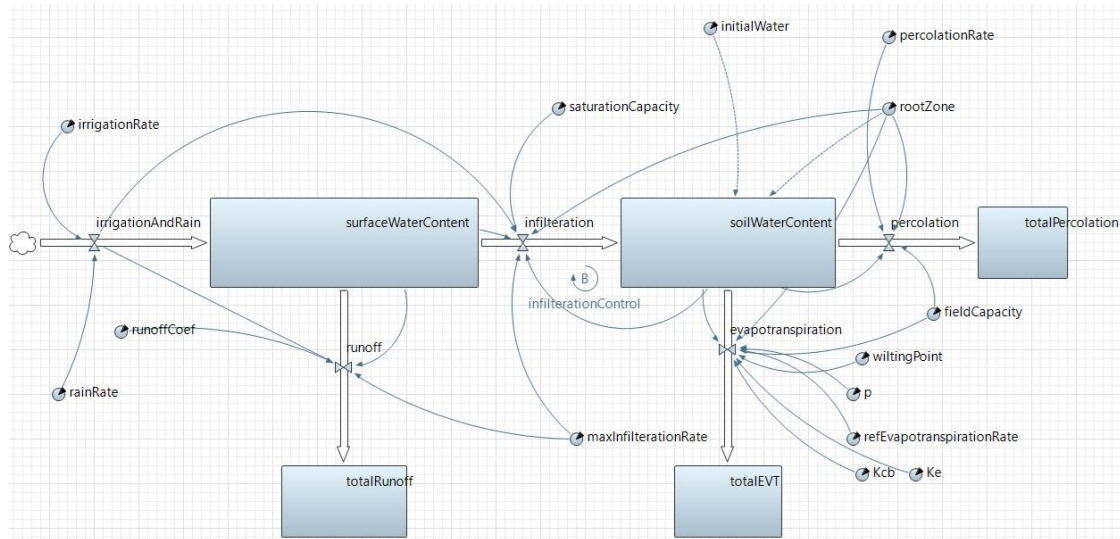

Fig. 6. System Dynamics stock and flow diagram for soil moisture dynamics model

### 5.2. System Architecture

After specifying the system's characteristics, it's time to design the system architecture. In this phase, various agent types need to be identified and ultimately, the overall structure of the system designed.

**Description of agents:**
- Central agent: This agent is the core of the system, controlling information flowing in the system, warnings and errors. This agent can transmit necessary information to other agents based on occurring events.
- Irrigation control agents: These agents have the decision-making ability for irrigation based on available data, deciding whether to irrigate or not.
- Sensors: These agents are responsible for data collection and, when necessary to share data, transmit it through the network.

**Systems overview:**
Considering the system's goals, agents' specifications, and their interactions, it's time to design the overall system structure. This system operates in an environment including:
- Time
- Weather conditions
- Soil properties
- Crop information

On the other hand, the following agents exist in the system:
- Central agent
- Irrigation control agents
- Sensors

The protocols available in the system are:
- Irrigation: Rules related to irrigation.
- Send notifications to the user: rules related to communication with the user.
- Measurement of environmental parameters: rules related to the measurement of parameters
- Error detection: Error detection rules.

And finally, the perceptions of the system are:

• Recognizing the need for irrigation: the ability of the system to process environmental information and recognize the need for irrigation.

• Necessary calculations: processing information related to plants and weather conditions for irrigation calculations.

• Taking inputs and processing them: processing input information including the type of irrigation method and schedules

• Water shortage detection: the system initially subtracts the amount of water irrigated from the initial water amount to calculate the amount of water that is currently available for irrigation, and if this amount is less than the amount of evaporation and transpiration in the next hours Water shortage is detected.

In this system, agents communicate each other and they control the irrigation. Based on the irrigation, soil moisture level changes and it should be considered in the environment modeling procedure. While the soil moisture changes, the sensors detect the change and they will transmit messages containing the collected data to the IoT network, and the next irrigation decision will be made based on these data. It is possible to draw the system overview diagram based on these definitions and details. Fig. 7 shows the notations in system overview diagram, and the Fig. 8 illustrates the overall system, agents, environment, interactions and database.

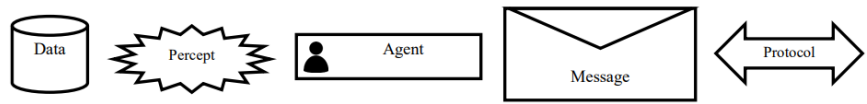

Fig. 7. Notations In System Overview Diagram

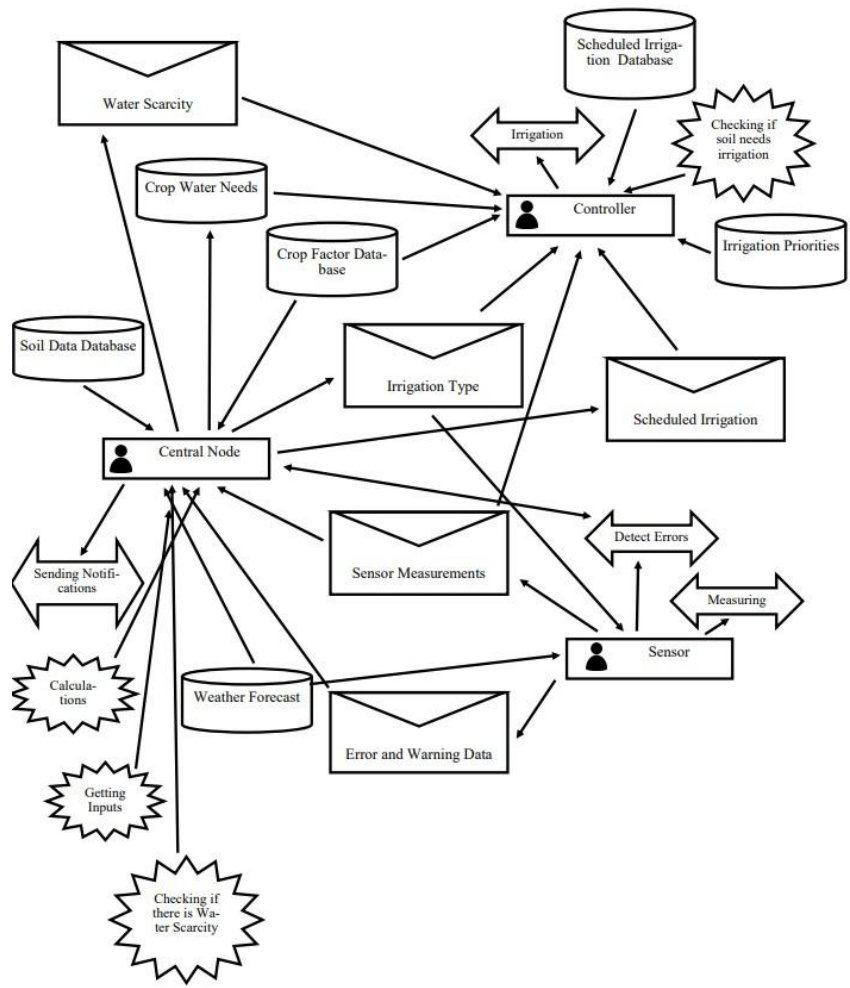

Fig. 8. System's overview

*5.3. System Details*

In the final phase of modeling using the Prometheus methodology, it's time to delve into the system's details. This phase includes the following steps:
• Define existing processes in the system
• Defining other specifications, including events and other parts of the system
• Precisely defining the behaviors of the agents

**This system includes the following processes:**
   • Environmental parameters measurement process: In this process, environmental parameters are measured by sensors and sent to the system. During this process, the system should be able to identify measurements that seem to be wrong (due to measurement errors) and inform the user in this case.
   • Automatic irrigation process in the presence of sufficient water: in this process, environmental parameters such as the rate of evaporation and transpiration, the predicted amount of rain, etc. According to the coefficients of the plant and its transpiration, irrigation is done if needed.
   • Automatic irrigation process in the presence of sufficient water: in this case, the calculations are the same as before and the only difference is that instead of watering as much as the plant needs, a part of this need is provided.
• Manual irrigation: In this case, the system only performs irrigation based on the schedule tables given to it by the user.

**Defining other system specifications:** One of the critical steps in the third phase of modeling using the Prometheus methodology is defining other system specifications. In this section, the events are introduced.

Events can happen at a time that may be specified or random, and depending on this event, the system or the environment will change. The events in this system are as follows:
   • Events related to climate change: including temperature changes, the amount of received ultraviolet rays, and rainfall. These values are measured in time cycles and are reported to the system if they change. Irrigation is done based on the amount of evapotranspiration calculated based on these factors.
   • Events related to changes in soil moisture: these changes are measured in specific time cycles and their value is reported to the system. If the soil moisture is less than the calculated limit, irrigation is done.
   • Events related to a critical decrease or increase in the amount of water in the soil: these events indicate an error in calculating the amount of water in the direction of irrigation and cause a change in the normal scenario of the system.
   • Events related to water shortage: situations in which the system detects water shortage and reduces the amount of irrigation according to the expert's opinion.
   • Events related to system failure: In these cases, the system has problems in performing its function and is forced to send a message to the user due to the change in the current scenario.

**Precisely defining the behaviors of the agents:**
Processes in this system are based on predefined rules. The proposed system performs its actions based on the real-time data measured by sensors or received from online weather forecast services. In this IoT-based system, different agents exhibit various behaviors, depending on specific events and scenarios. Also, some rules are defined to detect miscalculation or hardware malfunctions and notify the user. Each agent performs its action based on its current situation. These behaviors are listed in Table 3 as rules:

Table 3. Agent rules

| Rule number | Agent | Rule name | Scenario / Event | Rule description |
|---|---|---|---|---|
| 1 | Sensors | Measurement | Every time period | The sensor agent measures the environmental parameter. If the measurement error is high, the measurement can be done several times (for example, 40 times) and its average is considered as a suitable consolidation of the parameter. |
| 2 | Sensors | Sensor failure | The sensor does not receive information. | The agent sends the failure message to the central agent. |
| 3 | Sensors | Sensor error | The received information is of null data type, does not change, or is equal to the ceiling or floor of the sensor output values. It is assumed that the user calibrates the sensor in time, so there is no bias in the measurement and estimation of the desired parameter. | The agent sends the sensor error message to the central agent. |
| 4 | Sensors | Information sharing | Every time period | The sensor transmits the measured values to other agents. |
| 5 | Central agent | Request information from the server | Every time period | The system sends a request to the server to receive the required information. This information includes weather forecast information, soil information and plant texture. |
| 6 | Central agent | Error notification | When the central agent receives an error/warning | The system user will be notified of the error. |
| 7 | Central agent | Get information from the user | When the system is turned on | The information needed to configure. the system, such as the type of irrigation method, should be received from the user. |
| 8 | Central agent | Calculation and transfer of calculated data to the system | Every time period | The current evapotranspiration should be calculated and sent to the irrigation agents. |
| 9 | Central agent | Identify miscalculations | Every time period | It is checked whether soil moisture leaves the interval between wilting point and soil capacity or not. It is also checked that the total amount of irrigated water is close to the predicted value or differs by more than 20%. In those cases where the difference between the predicted value for irrigation and irrigated water is high, a possible miscalculation should be reported. |

| Rule number | Agent | Rule name | Scenario / Event | Rule description |
|---|---|---|---|---|
| 10 | Central agent | Identification of water shortage | Every time period | The amount of water that has been irrigated so far is subtracted from the total amount of initial water, to calculate the amount of remaining water. It is assumed that the amount of remaining water is not reduced by evaporation. In case of storing water in open tanks where there is also evaporation, the amount of evaporation should be reduced from the total amount of primary water. If the amount of remaining water is less than the evaporation and transpiration in the next hours, the water shortage is detected and the water shortage message is sent to the control agents. |
| 11 | Irrigation control | Irrigation in the early morning | Every early morning (exact time according to expert user's opinion) | Every morning, the soil moisture reaches the point determined by the expert so that there is no need to water during the hotter hours of the day. The determined point should be less than soil's capacity. |
| 12 | Irrigation control | Automatic irrigation control | Investigating the plant's need for irrigation | The control agent first checks whether the soil needs irrigation or not. If irrigation is needed, irrigation is done. If the amount of relative soil moisture in the future period, which is obtained by subtracting evaporation and transpiration during the period (which is calculated based on the current data and the forecasted data for the next hour) from the current soil moisture, was lower than the soil moisture threshold point, then irrigation It should be done and also if it was more than the capacity of the soil, even if it should have been irrigated according to the calculations, the irrigation should be stopped. It should be noted that the amount of irrigation should be equal to the expected amount of evaporation and transpiration.<br>The condition of irrigation is written as follows:<br><br>if ((soilWaterContent - periods*refEvapotranspirationRate*(Kcb + Ke)) <= (rootZone*threshold))<br><br>In this condition, soilWaterContent is the current moisture content of the soil determined by using the sensor, refEvapotranspirationRate is the reference evapotranspiration rate, kcb and ke are the evaporation and transpiration components of the crop coefficient, rootzone is the depth |

| Rule number | Agent | Rule name | Scenario / Event | Rule description |
|---|---|---|---|---|
| | | | | of the root zone and threshold is the humidity threshold. |
| 13 | Irrigation control | Stop irrigation during rainfall | Stop watering during rain | Stop irrigation during precipitation. |
| 14 | Irrigation control | Manual irrigation control | Irrigation based on the pre-defined schedule | The controlling agent should perform irrigation according to the table specified by the expert. |

## 6. Computer Simulation

One of the key benefits of creating agent-based models for IoT networks, is the ability to simulate the systems behavior. Sometimes it's necessary to create a simulation model to investigate the performance of the system. In our proposed model, it is possible to examine the system behavior with focus on irrigation control agent rules. Also, there is a possibility to model the irrigation system and its surrounding environment in simulation software. By existence of simulation software, like AnyLogic software, it is possible to create a copy of real-world systems in computers as a simulation model. It is recommended to run and examine the complex systems behavior on simulation software in order to improve the systems design. In addition to computer simulation, it is possible to use real-time data in order to monitor the implemented systems by running the agent-based model with real-time data.

AnyLogic software (AnyLogic 8.7.11 Professional) was used in order to model the system. Ease of modeling with this software, using JAVA programming language and possibility to combine agent-based and system dynamics approaches to create hybrid and dynamic models are the reasons that this simulation software was selected.

*6.1. Modeling the agents*

Three main agent types exist in proposed model. These agent types are:
- Central agent
- Controller agent
- Sensors

These agents are implemented with same structures, but behaving with different rules. The structure of each agent is similar to the Fig. 9.

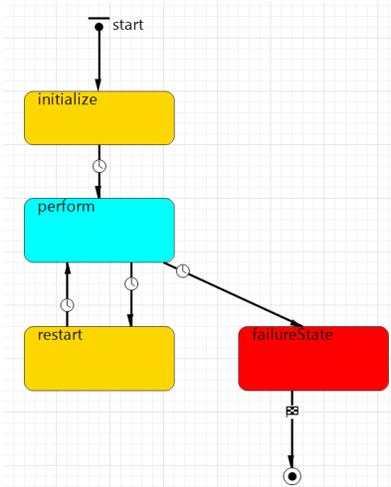

Fig. 9. Agent Structure in computer model

Each agent starts and goes to initialize state, then after a one-minute delay - which is considered due to the time that IoT nodes turn on and start to connect each other to shape the network – goes to the perform state and start its processes. In tasks state, in which agents do their tasks, each agent behaves based on its assigned rules. After fulfilling their tasks, each agent goes to deep sleep mode to reduce its energy consumption. After a determined time period, agents go to restart state. In this state they restart and after another one-minute delay they come back to task state in which start to fulfill their tasks. Please note each agent may also go to the failure state, by a determined probability. In fact, when a sensor or a chip fails to operate normally, it goes to the failure state.

*6.2. Examination of the soil moisture dynamics model*

In this section we test the SD model of soil moisture dynamics. The validation and verification of the SD model is important due to the reason that it models the environment of the agent-based model. The model must be a true model conceptually and also, be implemented true on simulation software. In order to investigate that the model concept have been truly created, the water movement and soil moisture dynamics was studied and the comments from experts was considered. In the case of the scenarios with true inputs such as evapotranspiration rate and infiltration rate, the model's functionality should be correct, Also the expected values of stocks, based on input and output rates were examined. These tests showed that the model was working correctly, if the input values were correct. In addition, the model's behavior on different conditions and extreme parameters was tested in order to examine its behavior. The information of model's behavior is listed in Table 4.

Table 4. Model examination

| Test | Parameter | Condition | Expected behavior | Is the simulated behavior similar to expected behavior? |
|---|---|---|---|---|
| 1 | Infiltration rate | The system is not irrigating and there is no ponded water on surface | Infiltration rate equals to zero | Yes |
| 2 | Infiltration rate | The system is irrigating and its rate is less than infiltration rate | Infiltration rate equals to irrigation rate | Yes |

| Test | Parameter | Condition | Expected behavior | Is the simulated behavior similar to expected behavior? |
|---|---|---|---|---|
| 3 | Infiltration rate | The system is irrigating and its rate is more than infiltration rate | Infiltration rate is at its maximum value | Yes |
| 4 | Infiltration rate | Soil moisture within the rootzone is saturated | There is no infiltration and water is ponded on the soil surface | Yes |
| 5 | Percolation flow | Soil moisture within the rootzone is more than fields capacity | Percolation flow is equal to percolation rate value | Yes |
| 6 | Percolation flow | Soil moisture within the rootzone is less than fields capacity | there is no percolation | Yes |
| 7 | Evapotranspiration rate | Soil moisture within the rootzone is less than fields capacity and more than threshold point | Evapotranspiration rate is equal to reference evapotranspiration rate multiplied by the crop factor | Yes |
| 8 | Ponded water | Irrigation rate is set on a great value | First, soil moisture increases to the saturation point. Then, the value of the ponded water increases linearly. | Yes |
| 9 | Soil moisture | Irrigation rate is set on a great value | First, soil moisture increases to the saturation point. Then, the soil moisture becomes constant. | Yes |
| 10 | Soil moisture | Irrigation rate is set on a zero (There is no irrigation) | Soil moisture decreases to wilting point (Due to the evapotranspiration). | Yes |
| 11 | Runoff | Irrigation rate is set on value exceeding the infiltration rate | Water exits the system as runoff, appropriate to the runoff coefficient | Yes |

| Test | Parameter | Condition | Expected behavior | Is the simulated behavior similar to expected behavior? |
|------|-----------|-----------|-------------------|-------------------------------------------------------|
| 12 | Evapotranspiration rate | Irrigation rate is set on zero, and soil moisture is decreased under the threshold value | The evapotranspiration rate decreases linearly and appropriate to the crop's stress coefficient | Yes |

Please note, this model operates truly in conditions that the assumptions are approximately true. For example, in situations that there is a significant surface water, the models cannot simulate the water surface evapotranspiration. However, the system in its automatic mode only operates in situations that the assumptions are correct.

*6.3. Examining the manual irrigation control rules by simulation*

After implementing the proposed system with an agent-based approach on AnyLogic software, irrigation control rules were examined by several experiments on simulation. Two types of irrigation control exist in proposed system. Scheduled control, which is used when user decides to plan the irrigation or at the cool hours after the sunrise, and the automated control.

The stoppage of the irrigation when it starts to rain is depicted in Fig. 10. Also, the system's scheduled irrigation rule is depicted on Fig. 11. The Fig. 12, shows the systems automatic irrigation rule for every morning.

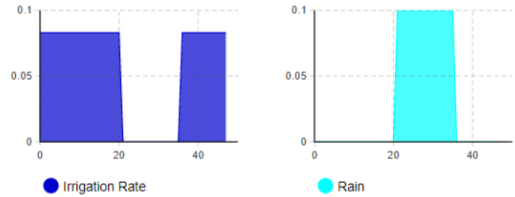

Fig. 10. Irrigation stoppage during rain

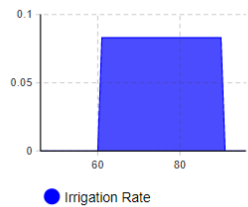

Fig. 11. Scheduled irrigation

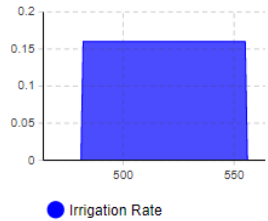

Fig. 12. Morning irrigation rule

*6.4. Examining the system's performance in automatic irrigation*

In this section, the automatic irrigation rule is examined. Fig. 13 shows the system performing irrigation rules in time periods.

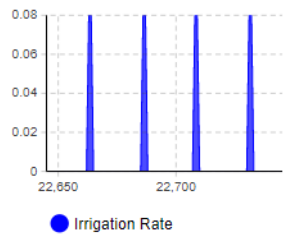

Fig. 13. The system performing automatic irrigation rules

In preferred model, several parameters exist. Many of these parameters are influential on the system's performance. These parameters may relate to the environment (like reference evapotranspiration rate), crops (crop factors), or the soil (like soil's infiltration rate). The mentioned parameters and their shorten names are listed in Table 5. These parameters can vary based on environmental factors like soil type. The proposed system must operate in different combinations of these parameters. While it is possible to test the model under a as there are 13 factors, it is time consuming to test the system's performance in all combinations of the parameters. Due to this limitation, a robust method was utilized to test the system. Some of the existing factors are controllable (like irrigation rate), and some of them are uncontrollable (like soil infiltration rate). 256-run resolution V fractional factorial designs with 13 factors was applied to the system to check its sensitivity to controllable and uncontrollable factors in three main types of soil. In fact, the model is tested based on parameters related to sandy, silty, and clay soil because all soil types are a mixture of these three main soil types. Moreover, resolution V design was chosen because it estimates all main effect and two factor interactions.

If the system operates acceptable in all combinations of the uncontrollable parameters in three main types of soils, it is possible to set the controllable parameters on appropriate values to optimize the system's performance.

Four responses were chosen for the test:
- The number of times that the soil moisture decreased to levels under the threshold point: It is unacceptable if the soil moisture decreases to levels under the threshold point
- Total percolated water: It is unacceptable if the system irrigates that much the water gets percolated to deeper levels of soil
- Total irrigated water: In optimal conditions, this response must be equal to the total evapotranspirated water
- Total time that the system was operating: This response is studied to minimize the system's operational time

Table 5. Factors and their assigned short names.

| Factor | Shorten Name |
|---|---|
| **Irrigation Rate** | rt |
| **Length of system's working periods** | tm |
| **Reference evapotranspiration rate** | evt |
| **Wilting point** | wp |
| **Field Capacity** | fc |
| **Soil Saturation** | st |
| **Percolation Rate** | pr |
| **Infiltration Rate** | nf |
| **Evaporation component of the Crop Factors** | ke |
| **Transpiration component of the Crop Factors** | kcb |
| **Root Zone Depth** | rz |
| **Fraction of total available soil moisture that exceeds the threshold point** | p |
| **Runoff coefficient** | ro |

The upper and lower levels of the factors in the experiments are listed in the Table 6. The root zone depth is based on the expert's opinion and considered for plants whose depth of the root zone is between 30 and 300 cm. The levels of the Length of system's working periods are experimentally and reasonably set to 20 minutes and 60 minutes. The lower level of the irrigation rate is set to half of the lower level of the maximum rate of infiltration rate of the soil and the upper rate is set to the lower level of this rate so that in different types of soil, it is possible to infiltrate this volume of water into the soil momentarily and surface water do not pool, because if pooling occurs, the runoff will increase water loss. The components of soil evaporation and plant transpiration were selected based on the available source [7], and reference evapotranspiration rates were chosen based on the approximate rate of reference evapotranspiration rate at temperatures of 10 and 40 degrees of Celsius and according to the Blaney-Criddle equation, which provides an approximation of reference evapotranspiration. Finally, infiltration rate, percolation rate, wilting point, the saturation capacity of the soil, field capacity, runoff coefficient and the fraction of total available soil moisture that exceeds the threshold point have been determined based on the data available in previous research [7, 35, 36, 39].

Table 6. Upper and Lower levels for parameters

| Parameters | Soil | | | | | |
|---|---|---|---|---|---|---|
| | **Sandy** | | **Silty** | | **Clay** | |
| | Upper level | Lower level | Upper level | Lower level | Upper level | Lower level |
| **evt (mm / minute)** | 0.0063 | 0.0021 | 0.0063 | 0.0021 | 0.0063 | 0.0021 |
| **rt (mm / minute)** | 0.334 | 0.167 | 0.126 | 0.063 | 0.025 | 0.0125 |
| **tm (minutes)** | 60 | 20 | 60 | 20 | 60 | 20 |
| **wp** | 0.1 | 0.05 | 0.22 | 0.07 | 0.24 | 0.17 |
| **fc** | 0.20 | 0.15 | 0.37 | 0.25 | 0.42 | 0.30 |

| Parameters | Soil | | | | | |
|---|---|---|---|---|---|---|
| | **Sandy** | | **Silty** | | **Clay** | |
| st | 0.45 | 0.4 | 0.5 | 0.45 | 0.51 | 0.4 |
| pr | 0.014 | 0.0035 | 0.0042 | 0.00083 | 0.0035 | 0.0007 |
| nf | 0.690 | 0.334 | 0.19 | 0.126 | 0.085 | 0.025 |
| kcb | 1.15 | 0.2 | 1.15 | 0.2 | 1.15 | 0.2 |
| ke | 0.3 | 0.1 | 0.3 | 0.1 | 0.3 | 0.1 |
| p | 0.65 | 0.2 | 0.65 | 0.2 | 0.65 | 0.2 |
| ro | 0.3 | 0.05 | 0.3 | 0.05 | 0.3 | 0.05 |
| rz (mm) | 3000 | 300 | 3000 | 300 | 3000 | 300 |

The 13 factors were also chosen to examine their effect on the system. These 13 factors were as follows:
The simulation had a warm-up about 2880 minutes (2 days) and the total run length of 17280 minutes (12 days), in order to get unbiased results.

The simulation result are as follows:
**Response 1: How many times the soil moisture decreased under the threshold point?**
Totally, the soil moisture never was decreased under the threshold point in all three types of the soil. This result shows that the system has a feasible solution.

**Response 2: The percolated water**
In all experiments and in all three soil types, no water was percolated to deeper levels of soil. This result shows that the system has maintained the soil moisture at the suitable interval between soils wilting point and the fields capacity to maintain moisture. Note that if the soil moisture reaches to higher amount in comparison to soil capacity, the crops health will be at risk.

**Response 3: Difference of toral irrigated water and the total evapotranspiration**
In an ideal test, the total irrigated water should be very close to total evapotranspiration. In all runs in three soil types, total irrigated water was very close to the total evapotranspiration. As the systems decision making is a Markovian decision-making process in which the system checks if the soil needs irrigation and irrigates if needed. In these experiments the systems irrigated water is very close to total evapotranspiration. The system's irrigation error is in automatic mode is presented in Table 7.

Table 7. Irrigation errors in automatic mode

| Error criteria | Soil type | | |
|---|---|---|---|
| | Sand | Silt | Clay |
| The sum of the squares of the difference between the amount of water required and the water irrigated (mm$^2$) | $1.837 \times 10^{-2}$ | $7.83 \times 10^{-3}$ | $9.85 \times 10^{-3}$ |
| The sum of the squares of the ratio of the difference between the required water and the irrigated water to the required water (no dimensions) | $2.159 \times 10^{-5}$ | $1.73 \times 10^{-6}$ | $1.90 \times 10^{-6}$ |

**Response 4: Total operating time of the system**
As one of the key barriers in designing an IoT system is their energy consumption, it is a valuable point in designing an IoT system to operate with minimum time, so the energy consumption will reduce. By analyzing the experiments, the effect of controllable factors and their interaction with uncontrollable factors should be determined and it is possible to set the controllable variables on an appropriate level.

To analyze the sensitivity of the total operating time response variable to other factors, the computer simulation model was tested on a resolution V 256-run $2^{13-5}$ design, and an analysis of variance was conducted. The analysis of variance results is available in the appendix A.

Based on the results from analysis of variance, the uncontrollable factors: reference evapotranspiration rate, the evaporation component of the crop factor and the transpiration component of the crop factor, and the controllable factors: working period length and irrigation rate are considered effective in all three types of sand.

In order to analyze the effects of the factors and their interactions, a response model was fitted on significant factors and interactions for each type of soil. The response models are available in the appendix B. Based on the results conducted from these experiments, it is possible to assess whether increasing the factor increases the response or vice versa. R-Squared and adjusted R-Squared measures were calculated and reported in Table 8. High R-Squared and adjusted R-Squared are showing the adequacy of the response surface models.

Table 8. $R^2$ and adjusted $R^2$ of response models

| Criteria | Soil | | |
|---|---|---|---|
|  | Sand | Silt | Clay |
| R-sq | 93.9% | 96.7% | 98.6% |
| adj R-sq | 93.5% | 96.5% | 98.5% |

Based on response models, the effect of the main 13 factors and their second level interactions were assessed in order to determine the level of controllable parameters. The contour and response surface plots of changes of system's operation time versus controllable factors in silty, sandy and clay soils are shown in the figures 14 to 19.

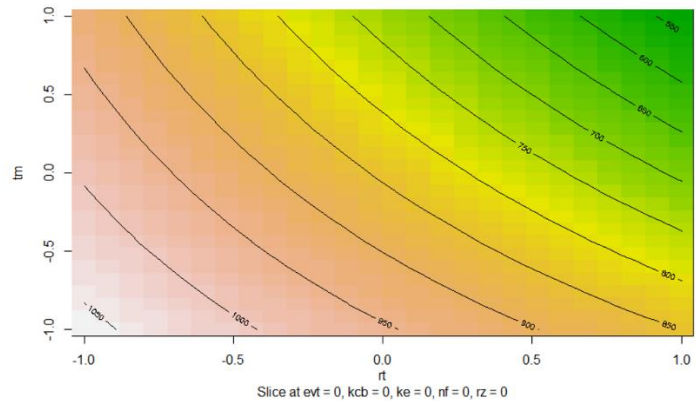

Fig. 14 Contour plot of changes in the level of system's operation time versus controllable factors (silty soil)

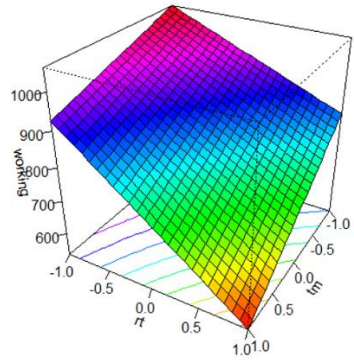

Fig. 15. Response surface plot of changes in the level of system's operation time versus controllable factors (silty soil)

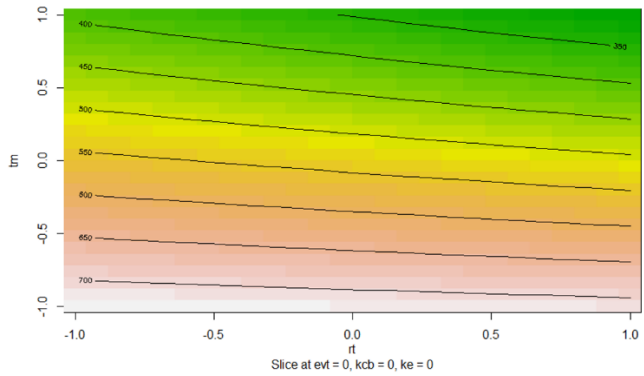

Fig. 16. Contour plot of changes in the level of system's operation time versus controllable factors (sandy soil)

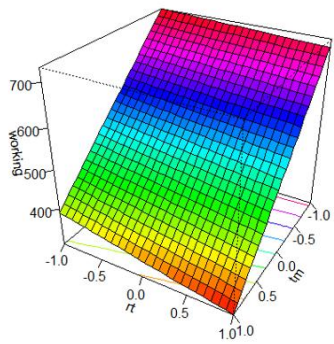

Fig. 17. Response surface plot of changes in the level of system's operation time versus controllable factors (sandy soil)

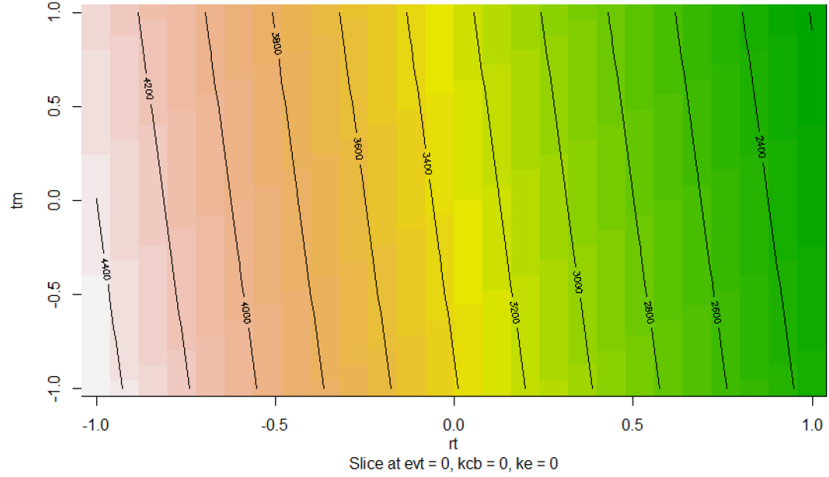

Fig. 18. Contour plot of changes in the level of system's operation time versus controllable factors (clay soil)

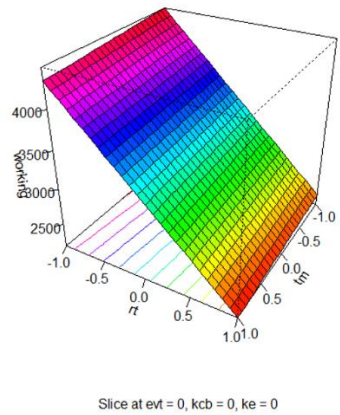

Fig. 19. Response surface plot of changes in the level of system's operation time versus controllable factors (clay soil)

In general, as expected, with the increase of the irrigation rate and the length of the working period length of the system, the total amount of working time of the device decreases and it is preferred to set these parameters on their upper levels. Positiveness and negativeness of the effects of the influential factors on system's working operational time is summarized in Table 9.

Table 9. The effective factors on system's working operational time

| Factor | Sandy soil | Silty soil | Clay soil |
|---|---|---|---|
| **rt** | - | - | - |
| **tm** | - | - | - |
| **evt** | + | + | + |
| **kcb** | + | + | + |
| **ke** | + | + | + |
| **evt✗rt** | - | - | - |

| Factor | Sandy soil | Silty soil | Clay soil |
|---|---|---|---|
| evt✖tm | + | + | |
| evt✖kcb | + | + | + |
| evt✖ke | + | + | + |
| rt✖tm | - | - | |
| rt✖kcb | - | - | - |
| tm✖kcb | + | + | |
| kcb✖ke | + | + | |

*6.5. Summary*

In this section, the designed system was examined in general. At first, the soil moisture dynamic sub-model was examined to validate and verify the model.

In the next step, the static rules of the system that regularly start or stop irrigation at the desired time were examined. It seems that these rules are correctly activated at the right time.

Finally, the automatic irrigation law was also examined. This rule was tested in 256 different runs and in three different soil types, and it performed well in all different combinations of high and low levels of controllable and uncontrollable factors. No amount of water was wasted in any case and the system never reached the soil moisture limit. On the other hand, the simulation results were also used to minimize the operating time of the device, and to achieve this goal, the irrigation rate and Length of the system's working periods should be set to their upper levels.

## 7. Conclusions

The primary objective of this study was to design a smart IoT-based irrigation system using an agent-based approach. The agent-based approach offers numerous benefits. Given the complexity of IoT networks, it is advisable to employ an appropriate approach for designing such systems.

Given the possibility of utilizing agent-oriented software engineering methodologies, we opted for the Prometheus methodology. Employing this methodology enables the structured design of an agent-based system. Also, an SD model was employed to represent the environment. This work presents the first instance of using a system dynamics approach to model the environment in the design of an irrigation system. By integrating SD with agent-based modeling, it became possible to analyze the mutual influences between the system and its surrounding environment.

To form the system's behavior, several functional rules were designed for different agent types. Following the design of the system's model, it was implemented in AnyLogic computer software to simulate the system and examine its operation. This hybrid model was validated and verified based on behavioral checks and was tested in several experimental runs.

The system's performance was tested in different conditions. In manual mode, the system was tested to ensure that it stops irrigation while raining and irrigates during the irrigation time windows. Also, the morning irrigation rule was tested. Moreover, the irrigation behavior in automatic irrigation mode was also tested by examining the effect of different factors on irrigation efficiency with a resolution V, $2^{13-5}$ experimental design in three main soil types.

According to the simulation results, the system effectively maintains soil moisture within the suitable range while minimizing percolated water. Additionally, response surface and contour plots were utilized to explore the impact of influential factors on the system's operational time. Based on these results, it is recommended that operating period times and irrigation rates be set at their highest possible levels to minimize the system's energy consumption.

The following are suggested for the future research:
- Considering the soil chemical balance in irrigation rules: This study managed the parameters to sustain effective moisture tension in the root zone, future activities could incorporate the control of soil chemical balance.
- Providing a robust method for prioritizing the crops for irrigation in water scarcity scenarios: Although the present research provided an optimized amount of irrigation, future research should provide a robust method for prioritization of the crops, to be irrigated in water deficit situations.

- Adding the capability of gathering data like plant images to check the plant's growth: This can be done by integrating Image Processing and machine learning that will help in assessing the plant health status through image processing [40].
- Utilizing the fuzzy sets to enhance the provided rules: Fuzzy sets can be used to improve the current rules to resolve the ambiguity of data [16].
- Utilizing Reinforcement learning to minimize systems operational time: In this article system performance was evaluated under several simulation runs; however, it is possible to use reinforcement learning to optimize the operation of the system and provide a comparison with the existing model.

**Declaration of generative AI and AI-assisted technologies in the writing process.**

During the preparation of this work the authors used Gemini and Chat GPT online services in order to check the writing and paraphrase the article's text. After using this tool/service, the authors reviewed and edited the content as needed and takes full responsibility for the content of the published article.

**Data availability**

Data will be made available on request.

**Funding Sources**

This research did not receive any specific grant from funding agencies in the public, commercial, or not-for-profit sectors.

**Appendix A:** The analysis of variance results of conducted experiments on system's operation time for different types of soil.

**The analysis of variance results of conducted experiments on system's operation time for sandy soil:**

|        | Df | Sum Sq  | Mean Sq | F value  | Pr(>F)      |     |
|--------|----|---------|---------|----------|-------------|-----|
| ro     | 1  | 0       | 0       | 0.000    | 1.000000    |     |
| evt    | 1  | 775280  | 775280  | 161.651  | < 2e-16     | *** |
| rt     | 1  | 177662  | 177662  | 37.044   | 7.90e-09    | *** |
| tm     | 1  | 8943090 | 8943090 | 1864.689 | < 2e-16     | *** |
| wp     | 1  | 0       | 0       | 0.000    | 1.000000    |     |
| fc     | 1  | 0       | 0       | 0.000    | 1.000000    |     |
| st     | 1  | 0       | 0       | 0.000    | 1.000000    |     |
| pr     | 1  | 0       | 0       | 0.000    | 1.000000    |     |
| nf     | 1  | 0       | 0       | 0.000    | 1.000000    |     |
| kcb    | 1  | 737022  | 737022  | 153.674  | < 2e-16     | *** |
| rz     | 1  | 0       | 0       | 0.000    | 1.000000    |     |
| ke     | 1  | 99540   | 99540   | 20.755   | 1.01e-05    | *** |
| p      | 1  | 0       | 0       | 0.000    | 1.000000    |     |
| ro:evt | 1  | 0       | 0       | 0.000    | 1.000000    |     |
| ro:rt  | 1  | 0       | 0       | 0.000    | 1.000000    |     |
| ro:tm  | 1  | 0       | 0       | 0.000    | 1.000000    |     |
| ro:wp  | 1  | 0       | 0       | 0.000    | 1.000000    |     |
| ro:fc  | 1  | 0       | 0       | 0.000    | 1.000000    |     |
| ro:st  | 1  | 0       | 0       | 0.000    | 1.000000    |     |
| ro:pr  | 1  | 0       | 0       | 0.000    | 1.000000    |     |

```
ro:nf      1      0       0    0.000 1.000000
ro:kcb     1      0       0    0.000 1.000000
ro:rz      1      0       0    0.000 1.000000
ro:ke      1      0       0    0.000 1.000000
ro:p       1      0       0    0.000 1.000000
evt:rt     1 154056  154056   32.122 6.37e-08 ***
evt:tm     1 584460  584460  121.864  < 2e-16 ***
evt:wp     1      0       0    0.000 1.000000
evt:fc     1      0       0    0.000 1.000000
evt:st     1      0       0    0.000 1.000000
evt:pr     1      0       0    0.000 1.000000
evt:nf     1      0       0    0.000 1.000000
evt:kcb    1 704760  704760  146.947  < 2e-16 ***
evt:rz     1      0       0    0.000 1.000000
evt:ke     1  72092   72092   15.032 0.000153 ***
evt:p      1      0       0    0.000 1.000000
rt:tm      1  74802   74802   15.597 0.000116 ***
rt:wp      1      0       0    0.000 1.000000
rt:fc      1      0       0    0.000 1.000000
rt:st      1      0       0    0.000 1.000000
rt:pr      1      0       0    0.000 1.000000
rt:nf      1      0       0    0.000 1.000000
rt:kcb     1 138756  138756   28.932 2.55e-07 ***
rt:rz      1      0       0    0.000 1.000000
```

| | | | | | |
|---|---|---|---|---|---|
| rt:ke  | 1 | 45582 | 45582 | 9.504 | 0.002406 ** |
| rt:p   | 1 | 0 | 0 | 0.000 | 1.000000 |
| tm:wp  | 1 | 0 | 0 | 0.000 | 1.000000 |
| tm:fc  | 1 | 0 | 0 | 0.000 | 1.000000 |
| tm:st  | 1 | 0 | 0 | 0.000 | 1.000000 |
| tm:pr  | 1 | 0 | 0 | 0.000 | 1.000000 |
| tm:nf  | 1 | 0 | 0 | 0.000 | 1.000000 |
| tm:kcb | 1 | 557262 | 557262 | 116.193 | < 2e-16 *** |
| tm:rz  | 1 | 0 | 0 | 0.000 | 1.000000 |
| tm:ke  | 1 | 15750 | 15750 | 3.284 | 0.071786 . |
| tm:p   | 1 | 0 | 0 | 0.000 | 1.000000 |
| wp:fc  | 1 | 0 | 0 | 0.000 | 1.000000 |
| wp:st  | 1 | 0 | 0 | 0.000 | 1.000000 |
| wp:pr  | 1 | 0 | 0 | 0.000 | 1.000000 |
| wp:nf  | 1 | 0 | 0 | 0.000 | 1.000000 |
| wp:kcb | 1 | 0 | 0 | 0.000 | 1.000000 |
| wp:rz  | 1 | 0 | 0 | 0.000 | 1.000000 |
| wp:ke  | 1 | 0 | 0 | 0.000 | 1.000000 |
| wp:p   | 1 | 0 | 0 | 0.000 | 1.000000 |
| fc:st  | 1 | 0 | 0 | 0.000 | 1.000000 |
| fc:pr  | 1 | 0 | 0 | 0.000 | 1.000000 |
| fc:nf  | 1 | 0 | 0 | 0.000 | 1.000000 |
| fc:kcb | 1 | 0 | 0 | 0.000 | 1.000000 |
| fc:rz  | 1 | 0 | 0 | 0.000 | 1.000000 |
| fc:ke | 1 | 0 | 0 | 0.000 | 1.000000 |
| fc:p | 1 | 56 | 56 | 0.012 | 0.913892 |
| st:pr | 1 | 0 | 0 | 0.000 | 1.000000 |
| st:nf | 1 | 0 | 0 | 0.000 | 1.000000 |
| st:kcb | 1 | 0 | 0 | 0.000 | 1.000000 |
| st:rz | 1 | 0 | 0 | 0.000 | 1.000000 |
| st:ke | 1 | 0 | 0 | 0.000 | 1.000000 |
| st:p | 1 | 0 | 0 | 0.000 | 1.000000 |
| pr:nf | 1 | 0 | 0 | 0.000 | 1.000000 |
| pr:kcb | 1 | 0 | 0 | 0.000 | 1.000000 |
| pr:rz | 1 | 0 | 0 | 0.000 | 1.000000 |
| pr:ke | 1 | 0 | 0 | 0.000 | 1.000000 |
| pr:p | 1 | 0 | 0 | 0.000 | 1.000000 |
| nf:kcb | 1 | 0 | 0 | 0.000 | 1.000000 |
| nf:rz | 1 | 8742 | 8742 | 1.823 | 0.178839 |
| nf:ke | 1 | 0 | 0 | 0.000 | 1.000000 |
| nf:p | 1 | 0 | 0 | 0.000 | 1.000000 |
| kcb:rz | 1 | 0 | 0 | 0.000 | 1.000000 |
| kcb:ke | 1 | 66822 | 66822 | 13.933 | 0.000261 *** |
| kcb:p | 1 | 0 | 0 | 0.000 | 1.000000 |
| rz:ke | 1 | 0 | 0 | 0.000 | 1.000000 |
| rz:p | 1 | 0 | 0 | 0.000 | 1.000000 |
| ke:p | 1 | 0 | 0 | 0.000 | 1.000000 |
| Residuals | 164 | 786548 | 4796 | | |

---

Signif. codes:  0 '***' 0.001 '**' 0.01 '*' 0.05 '.' 0.1 ' ' 1

**The analysis of variance results of conducted experiments on system's operation time for silty soil:**

|       | Df | Sum Sq | Mean Sq | F value | Pr(>F) |
|-------|----|--------|---------|---------|--------|
| ro    | 1  | 0        | 0        | 0.000     | 0.99959 |
| evt   | 1  | 21581251 | 21581251 | 1493.514  | < 2e-16 *** |
| rt    | 1  | 5883353  | 5883353  | 407.153   | < 2e-16 *** |
| tm    | 1  | 3231231  | 3231231  | 223.615   | < 2e-16 *** |
| wp    | 1  | 0        | 0        | 0.000     | 0.99959 |
| fc    | 1  | 0        | 0        | 0.000     | 0.99959 |
| st    | 1  | 0        | 0        | 0.000     | 0.99959 |
| pr    | 1  | 0        | 0        | 0.000     | 0.99959 |
| nf    | 1  | 0        | 0        | 0.000     | 0.99959 |
| kcb   | 1  | 22103515 | 22103515 | 1529.657  | < 2e-16 *** |
| rz    | 1  | 0        | 0        | 0.000     | 0.99959 |
| ke    | 1  | 681347   | 681347   | 47.152    | 1.29e-10 *** |
| p     | 1  | 0        | 0        | 0.000     | 0.99959 |
| ro:evt | 1 | 0        | 0        | 0.000     | 0.99959 |
| ro:rt  | 1 | 0        | 0        | 0.000     | 0.99959 |
| ro:tm  | 1 | 0        | 0        | 0.000     | 0.99959 |
| ro:wp  | 1 | 0        | 0        | 0.000     | 0.99959 |
| ro:fc  | 1 | 0        | 0        | 0.000     | 0.99959 |
| ro:st  | 1 | 0        | 0        | 0.000     | 0.99959 |
| ro:pr | 1 | 0 | 0 | 0.000 | 0.99959 |
| ro:nf | 1 | 0 | 0 | 0.000 | 0.99959 |
| ro:kcb | 1 | 0 | 0 | 0.000 | 0.99959 |
| ro:rz | 1 | 0 | 0 | 0.000 | 0.99959 |
| ro:ke | 1 | 0 | 0 | 0.000 | 0.99959 |
| ro:p | 1 | 0 | 0 | 0.000 | 0.99959 |
| evt:rt | 1 | 3498536 | 3498536 | 242.114 | < 2e-16 *** |
| evt:tm | 1 | 450996 | 450996 | 31.211 | 9.44e-08 *** |
| evt:wp | 1 | 0 | 0 | 0.000 | 0.99959 |
| evt:fc | 1 | 0 | 0 | 0.000 | 0.99959 |
| evt:st | 1 | 0 | 0 | 0.000 | 0.99959 |
| evt:pr | 1 | 0 | 0 | 0.000 | 0.99959 |
| evt:nf | 1 | 0 | 0 | 0.000 | 0.99959 |
| evt:kcb | 1 | 12114315 | 12114315 | 838.362 | < 2e-16 *** |
| evt:rz | 1 | 0 | 0 | 0.000 | 0.99959 |
| evt:ke | 1 | 485199 | 485199 | 33.578 | 3.42e-08 *** |
| evt:p | 1 | 0 | 0 | 0.000 | 0.99959 |
| rt:tm | 1 | 532261 | 532261 | 36.835 | 8.62e-09 *** |
| rt:wp | 1 | 0 | 0 | 0.000 | 0.99959 |
| rt:fc | 1 | 0 | 0 | 0.000 | 0.99959 |
| rt:st | 1 | 0 | 0 | 0.000 | 0.99959 |
| rt:pr | 1 | 0 | 0 | 0.000 | 0.99959 |
| rt:nf | 1 | 0 | 0 | 0.000 | 0.99959 |
| rt:kcb | 1 | 2960335 | 2960335 | 204.868 | < 2e-16 *** |
| rt:rz | 1 | 0 | 0 | 0.000 | 0.99959 |
| rt:ke | 1 | 41846 | 41846 | 2.896 | 0.09070 . |
| rt:p | 1 | 0 | 0 | 0.000 | 0.99959 |
| tm:wp | 1 | 0 | 0 | 0.000 | 0.99959 |
| tm:fc | 1 | 0 | 0 | 0.000 | 0.99959 |
| tm:st | 1 | 0 | 0 | 0.000 | 0.99959 |
| tm:pr | 1 | 0 | 0 | 0.000 | 0.99959 |
| tm:nf | 1 | 0 | 0 | 0.000 | 0.99959 |
| tm:kcb | 1 | 724946 | 724946 | 50.169 | 3.94e-11 *** |
| tm:rz | 1 | 0 | 0 | 0.000 | 0.99959 |
| tm:ke | 1 | 1256 | 1256 | 0.087 | 0.76852 |
| tm:p | 1 | 0 | 0 | 0.000 | 0.99959 |
| wp:fc | 1 | 0 | 0 | 0.000 | 0.99959 |
| wp:st | 1 | 0 | 0 | 0.000 | 0.99959 |
| wp:pr | 1 | 0 | 0 | 0.000 | 0.99959 |
| wp:nf | 1 | 0 | 0 | 0.000 | 0.99959 |
| wp:kcb | 1 | 0 | 0 | 0.000 | 0.99959 |
| wp:rz | 1 | 0 | 0 | 0.000 | 0.99959 |
| wp:ke | 1 | 0 | 0 | 0.000 | 0.99959 |
| wp:p | 1 | 0 | 0 | 0.000 | 0.99959 |
| fc:st | 1 | 0 | 0 | 0.000 | 0.99959 |
| fc:pr | 1 | 0 | 0 | 0.000 | 0.99959 |
| fc:nf | 1 | 0 | 0 | 0.000 | 0.99959 |
| fc:kcb | 1 | 0 | 0 | 0.000 | 0.99959 |
| fc:rz   | 1 | 0 | 0 | 0.000 | 0.99959 |
| fc:ke   | 1 | 0 | 0 | 0.000 | 0.99959 |
| fc:p    | 1 | 304 | 304 | 0.021 | 0.88484 |
| st:pr   | 1 | 0 | 0 | 0.000 | 0.99959 |
| st:nf   | 1 | 0 | 0 | 0.000 | 0.99959 |
| st:kcb  | 1 | 0 | 0 | 0.000 | 0.99959 |
| st:rz   | 1 | 0 | 0 | 0.000 | 0.99959 |
| st:ke   | 1 | 0 | 0 | 0.000 | 0.99959 |
| st:p    | 1 | 0 | 0 | 0.000 | 0.99959 |
| pr:nf   | 1 | 0 | 0 | 0.000 | 0.99959 |
| pr:kcb  | 1 | 0 | 0 | 0.000 | 0.99959 |
| pr:rz   | 1 | 0 | 0 | 0.000 | 0.99959 |
| pr:ke   | 1 | 0 | 0 | 0.000 | 0.99959 |
| pr:p    | 1 | 0 | 0 | 0.000 | 0.99959 |
| nf:kcb  | 1 | 0 | 0 | 0.000 | 0.99959 |
| nf:rz   | 1 | 106561 | 106561 | 7.375 | 0.00732 ** |
| nf:ke   | 1 | 0 | 0 | 0.000 | 0.99959 |
| nf:p    | 1 | 0 | 0 | 0.000 | 0.99959 |
| kcb:rz  | 1 | 0 | 0 | 0.000 | 0.99959 |
| kcb:ke  | 1 | 161956 | 161956 | 11.208 | 0.00101 ** |
| kcb:p   | 1 | 0 | 0 | 0.000 | 0.99959 |
| rz:ke   | 1 | 0 | 0 | 0.000 | 0.99959 |
| rz:p    | 1 | 0 | 0 | 0.000 | 0.99959 |
| ke:p    | 1 | 0 | 0 | 0.000 | 0.99959 |

Residuals  164  2369796    14450

---

Signif. codes:  0 '***' 0.001 '**' 0.01 '*' 0.05 '.' 0.1 ' ' 1

**The analysis of variance results of conducted experiments on system's operation time for clay soil:**

|      | Df | Sum Sq    | Mean Sq   | F value  | Pr(>F)   |    |
|------|----|-----------|-----------|----------|----------|----|
| ro   | 1  | 0         | 0         | 0.000    | 1.00000  |    |
| evt  | 1  | 631843632 | 631843632 | 4515.313 | < 2e-16  | ***|
| rt   | 1  | 290753652 | 290753652 | 2077.798 | < 2e-16  | ***|
| tm   | 1  | 1479872   | 1479872   | 10.576   | 0.00139  | ** |
| wp   | 1  | 0         | 0         | 0.000    | 1.00000  |    |
| fc   | 1  | 0         | 0         | 0.000    | 1.00000  |    |
| st   | 1  | 0         | 0         | 0.000    | 1.00000  |    |
| pr   | 1  | 0         | 0         | 0.000    | 1.00000  |    |
| nf   | 1  | 0         | 0         | 0.000    | 1.00000  |    |
| kcb  | 1  | 753639756 | 753639756 | 5385.699 | < 2e-16  | ***|
| rz   | 1  | 0         | 0         | 0.000    | 1.00000  |    |
| ke   | 1  | 31488932  | 31488932  | 225.028  | < 2e-16  | ***|
| p    | 1  | 0         | 0         | 0.000    | 1.00000  |    |
| ro:evt | 1 | 0        | 0         | 0.000    | 1.00000  |    |
| ro:rt  | 1 | 0        | 0         | 0.000    | 1.00000  |    |
| ro:tm  | 1 | 0        | 0         | 0.000    | 1.00000  |    |
| ro:wp  | 1 | 0        | 0         | 0.000    | 1.00000  |    |
| ro:fc  | 1 | 0        | 0         | 0.000    | 1.00000  |    |

| | | | | | |
|---|---|---|---|---|---|
| ro:st | 1 | 0 | 0 | 0.000 | 1.00000 |
| ro:pr | 1 | 0 | 0 | 0.000 | 1.00000 |
| ro:nf | 1 | 0 | 0 | 0.000 | 1.00000 |
| ro:kcb | 1 | 0 | 0 | 0.000 | 1.00000 |
| ro:rz | 1 | 0 | 0 | 0.000 | 1.00000 |
| ro:ke | 1 | 0 | 0 | 0.000 | 1.00000 |
| ro:p | 1 | 0 | 0 | 0.000 | 1.00000 |
| evt:rt | 1 | 71681622 | 71681622 | 512.255 | < 2e-16 *** |
| evt:tm | 1 | 10920 | 10920 | 0.078 | 0.78032 |
| evt:wp | 1 | 0 | 0 | 0.000 | 1.00000 |
| evt:fc | 1 | 0 | 0 | 0.000 | 1.00000 |
| evt:st | 1 | 0 | 0 | 0.000 | 1.00000 |
| evt:pr | 1 | 0 | 0 | 0.000 | 1.00000 |
| evt:nf | 1 | 0 | 0 | 0.000 | 1.00000 |
| evt:kcb | 1 | 185272932 | 185272932 | 1324.007 | < 2e-16 *** |
| evt:rz | 1 | 0 | 0 | 0.000 | 1.00000 |
| evt:ke | 1 | 6741812 | 6741812 | 48.179 | 8.61e-11 *** |
| evt:p | 1 | 0 | 0 | 0.000 | 1.00000 |
| rt:tm | 1 | 930 | 930 | 0.007 | 0.93512 |
| rt:wp | 1 | 0 | 0 | 0.000 | 1.00000 |
| rt:fc | 1 | 0 | 0 | 0.000 | 1.00000 |
| rt:st | 1 | 0 | 0 | 0.000 | 1.00000 |
| rt:pr | 1 | 0 | 0 | 0.000 | 1.00000 |
| rt:nf | 1 | 0 | 0 | 0.000 | 1.00000 |
| rt:kcb | 1 | 82637190 | 82637190 | 590.546 | < 2e-16 *** |
| rt:rz  | 1 | 0 | 0 | 0.000 | 1.00000 |
| rt:ke  | 1 | 5757600 | 5757600 | 41.145 | 1.45e-09 *** |
| rt:p   | 1 | 0 | 0 | 0.000 | 1.00000 |
| tm:wp  | 1 | 0 | 0 | 0.000 | 1.00000 |
| tm:fc  | 1 | 0 | 0 | 0.000 | 1.00000 |
| tm:st  | 1 | 0 | 0 | 0.000 | 1.00000 |
| tm:pr  | 1 | 0 | 0 | 0.000 | 1.00000 |
| tm:nf  | 1 | 0 | 0 | 0.000 | 1.00000 |
| tm:kcb | 1 | 10100 | 10100 | 0.072 | 0.78853 |
| tm:rz  | 1 | 0 | 0 | 0.000 | 1.00000 |
| tm:ke  | 1 | 57360 | 57360 | 0.410 | 0.52291 |
| tm:p   | 1 | 0 | 0 | 0.000 | 1.00000 |
| wp:fc  | 1 | 0 | 0 | 0.000 | 1.00000 |
| wp:st  | 1 | 0 | 0 | 0.000 | 1.00000 |
| wp:pr  | 1 | 0 | 0 | 0.000 | 1.00000 |
| wp:nf  | 1 | 0 | 0 | 0.000 | 1.00000 |
| wp:kcb | 1 | 0 | 0 | 0.000 | 1.00000 |
| wp:rz  | 1 | 0 | 0 | 0.000 | 1.00000 |
| wp:ke  | 1 | 0 | 0 | 0.000 | 1.00000 |
| wp:p   | 1 | 0 | 0 | 0.000 | 1.00000 |
| fc:st  | 1 | 0 | 0 | 0.000 | 1.00000 |
| fc:pr  | 1 | 0 | 0 | 0.000 | 1.00000 |
| fc:nf  | 1 | 0 | 0 | 0.000 | 1.00000 |
| fc:kcb | 1 | 0 | 0 | 0.000 | 1.00000 |
| fc:rz | 1 | 0 | 0 | 0.000 | 1.00000 |
| fc:ke | 1 | 0 | 0 | 0.000 | 1.00000 |
| fc:p | 1 | 346332 | 346332 | 2.475 | 0.11760 |
| st:pr | 1 | 0 | 0 | 0.000 | 1.00000 |
| st:nf | 1 | 0 | 0 | 0.000 | 1.00000 |
| st:kcb | 1 | 0 | 0 | 0.000 | 1.00000 |
| st:rz | 1 | 0 | 0 | 0.000 | 1.00000 |
| st:ke | 1 | 0 | 0 | 0.000 | 1.00000 |
| st:p | 1 | 0 | 0 | 0.000 | 1.00000 |
| pr:nf | 1 | 0 | 0 | 0.000 | 1.00000 |
| pr:kcb | 1 | 0 | 0 | 0.000 | 1.00000 |
| pr:rz | 1 | 0 | 0 | 0.000 | 1.00000 |
| pr:ke | 1 | 0 | 0 | 0.000 | 1.00000 |
| pr:p | 1 | 0 | 0 | 0.000 | 1.00000 |
| nf:kcb | 1 | 0 | 0 | 0.000 | 1.00000 |
| nf:rz | 1 | 20592 | 20592 | 0.147 | 0.70176 |
| nf:ke | 1 | 0 | 0 | 0.000 | 1.00000 |
| nf:p | 1 | 0 | 0 | 0.000 | 1.00000 |
| kcb:rz | 1 | 0 | 0 | 0.000 | 1.00000 |
| kcb:ke | 1 | 4290 | 4290 | 0.031 | 0.86122 |
| kcb:p | 1 | 0 | 0 | 0.000 | 1.00000 |
| rz:ke | 1 | 0 | 0 | 0.000 | 1.00000 |
| rz:p | 1 | 0 | 0 | 0.000 | 1.00000 |

ke:p         1     0      0   0.000  1.00000

Residuals  164  22949098   139934

---

Signif. codes:  0 '***' 0.001 '**' 0.01 '*' 0.05 '.' 0.1 ' ' 1

**Appendix B:** The response surface model of system's operation time for different types of soil.

**The linearized response model of system's operation time for sandy soil and normal probability plot of its residuals:**

Residuals:

| Min | 1Q | Median | 3Q | Max |
|---|---|---|---|---|
| -106.625 | -40.469 | 6.438 | 36.875 | 147.813 |

Coefficients:

|  | Estimate | Std. Error | t value | Pr(>|t|) |  |
|---|---|---|---|---|---|
| (Intercept) | 534.594 | 3.719 | 143.762 | < 2e-16 | *** |
| evt | 55.031 | 3.719 | 14.799 | < 2e-16 | *** |
| kcb | 53.656 | 3.719 | 14.429 | < 2e-16 | *** |
| ke | 19.719 | 3.719 | 5.303 | 2.57e-07 | *** |
| rt | -26.344 | 3.719 | -7.084 | 1.52e-11 | *** |
| tm | -186.906 | 3.719 | -50.262 | < 2e-16 | *** |
| evt:tm | 47.781 | 3.719 | 12.849 | < 2e-16 | *** |
| rt:tm | -17.094 | 3.719 | -4.597 | 6.91e-06 | *** |
| evt:rt | -24.531 | 3.719 | -6.597 | 2.62e-10 | *** |
| evt:kcb | 52.469 | 3.719 | 14.110 | < 2e-16 | *** |

| | | | | |
|---|---|---|---|---|
| evt:ke | 16.781 | 3.719 | 4.513 | 9.99e-06 *** |
| kcb:rt | -23.281 | 3.719 | -6.261 | 1.73e-09 *** |
| kcb:tm | 46.656 | 3.719 | 12.547 | < 2e-16 *** |
| kcb:ke | 16.156 | 3.719 | 4.345 | 2.05e-05 *** |

---

Signif. codes:  0 '***' 0.001 '**' 0.01 '*' 0.05 '.' 0.1 ' ' 1

Residual standard error: 59.5 on 242 degrees of freedom

Multiple R-squared:  0.9386,     Adjusted R-squared:  0.9353

F-statistic: 284.3 on 13 and 242 DF,  p-value: < 2.2e-16

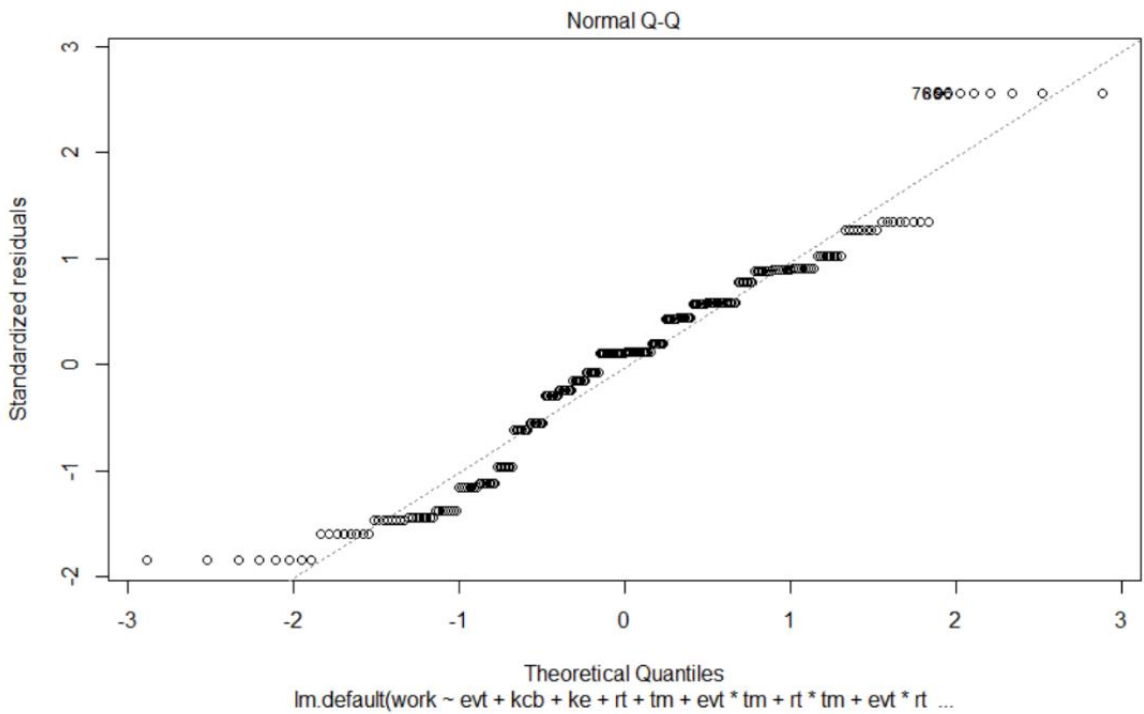

Fig. B.1. Normal probability plot of residuals of the linearized response model of system's operation time for sandy soil

**The linearized response model of system's operation time for silty soil and normal probability plot of its residuals:**

Residuals:

```
   Min      1Q  Median      3Q     Max
215.500  65.928  0.973-  71.857- 211.109-
```

Coefficients:

```
              Estimate Std. Error  t value  Pr(>|t|)
(Intercept)    842.902      6.378  132.168  < 2e-16 ***
evt            290.348      6.378   45.527  < 2e-16 ***
kcb            293.840      6.378   46.074  < 2e-16 ***
ke              51.590      6.378    8.089 2.90e-14 ***
rt            -151.598      6.378  -23.771  < 2e-16 ***
tm            -112.348      6.378  -17.616  < 2e-16 ***
evt:tm          41.973      6.378    6.581 2.86e-10 ***
rt:tm          -45.598      6.378   -7.150 1.02e-11 ***
evt:rt        -116.902      6.378  -18.330  < 2e-16 ***
evt:kcb        217.535      6.378   34.110  < 2e-16 ***
evt:ke          43.535      6.378    6.826 6.96e-11 ***
kcb:rt        -107.535      6.378  -16.862  < 2e-16 ***
kcb:tm          53.215      6.378    8.344 5.51e-15 ***
kcb:ke          25.152      6.378    3.944 0.000105 ***
```

---

Signif. codes:  0 '***' 0.001 '**' 0.01 '*' 0.05 '.' 0.1 ' ' 1

Residual standard error: 102 on 242 degrees of freedom

Multiple R-squared: 0.9672, Adjusted R-squared: 0.9655

F-statistic: 549.7 on 13 and 242 DF, p-value: < 2.2e-16

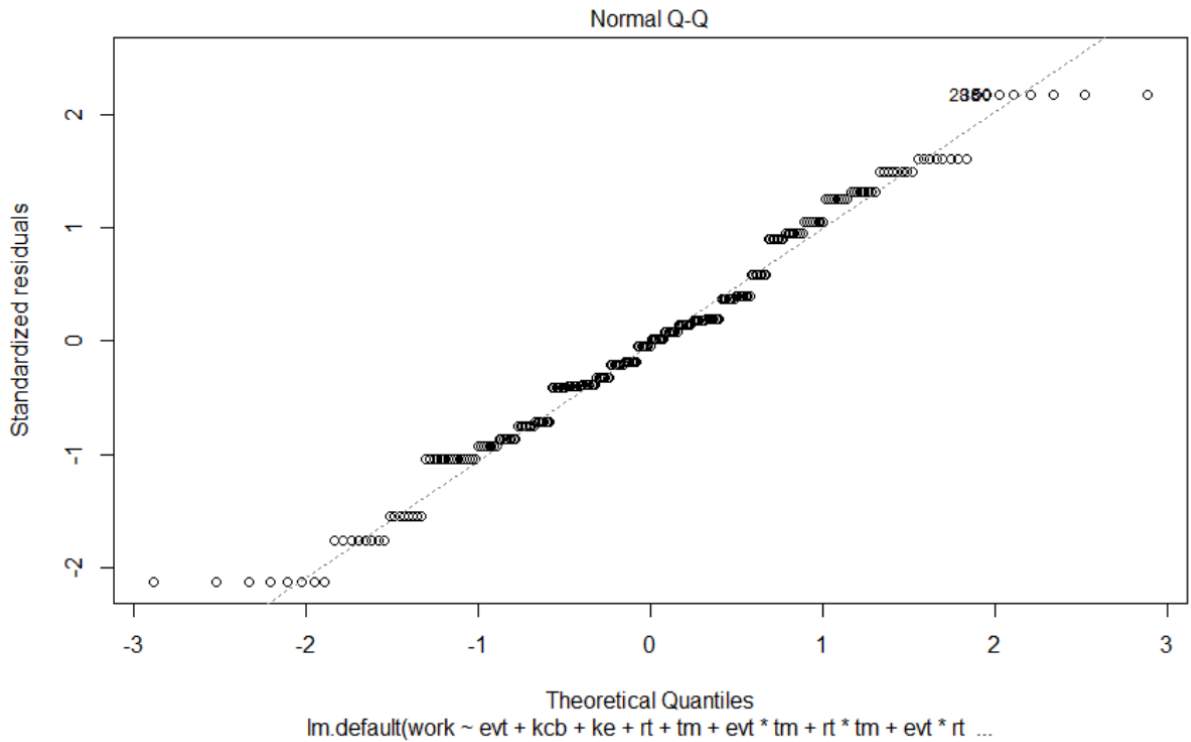

Fig. B.2. Normal probability plot of residuals of the linearized response model of system's operation time for silty soil

**The linearized response model of system's operation time for clay soil and normal probability plot of its residuals:**

Residuals:

   Min    1Q  Median    3Q    Max

-533.06 -266.59 -17.87 255.91 633.31

Coefficients:

       Estimate Std. Error t value Pr(>|t|)

```
(Intercept)  3334.84    21.52 154.985  < 2e-16 ***
evt          1571.03    21.52  73.013  < 2e-16 ***
kcb          1715.78    21.52  79.740  < 2e-16 ***
ke            350.72    21.52  16.299  < 2e-16 ***
rt          -1065.72    21.52 -49.529  < 2e-16 ***
tm            -76.03    21.52  -3.534  0.00049 ***
evt:rt       -529.16    21.52 -24.592  < 2e-16 ***
evt:kcb       850.72    21.52  39.537  < 2e-16 ***
evt:ke        162.28    21.52   7.542 8.92e-13 ***
kcb:rt       -568.16    21.52 -26.405  < 2e-16 ***
---
```

Signif. codes:  0 '***' 0.001 '**' 0.01 '*' 0.05 '.' 0.1 ' ' 1

Residual standard error: 344.3 on 246 degrees of freedom

Multiple R-squared: 0.986,    Adjusted R-squared: 0.9855

F-statistic: 1927 on 9 and 246 DF,  p-value: < 2.2e-16

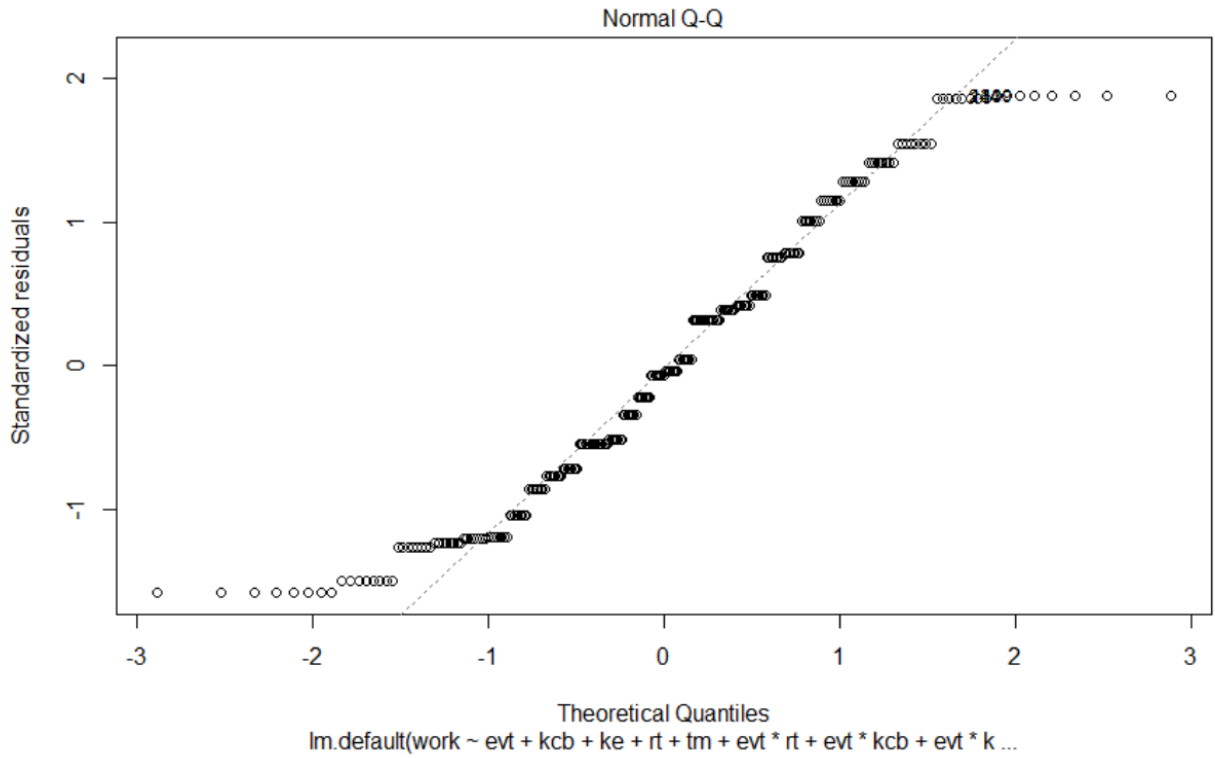

Fig. B.3. Normal probability plot of residuals of the linearized response model of system's operation time for clay soil